\documentclass[useAMS,usenatbib]{mn2e}
\usepackage{float}
\usepackage{graphicx}
\usepackage{amssymb}
\usepackage{amsmath}
\usepackage{datetime}
\usepackage[normalem]{ulem}
\usepackage{times} 
\usepackage[export]{adjustbox}
\usepackage{soul}

\usepackage{color}
\usepackage{soul}
\definecolor{fabian}{rgb}{1,0,1}

\definecolor{stan}{rgb}{0,0,1}

\definecolor{ryo}{rgb}{1,0,0}

\newcommand{\Msun}{{\rm M}_\odot}
\newcommand{\Rsun}{{\rm R}_\odot}
\newcommand{\Rstar}{R_\ast}
\newcommand{\Mstar}{M_\ast}

\newcommand{\Ebind}{{E_\ast}}
\newcommand{\DMtot}{\Delta M}
\newcommand{\DKEtot}{\Delta K}
\newcommand{\BEfin}{E_{\rm fin}}

\newcommand{\Lsun}{{\rm L}_\odot}

\newcommand{\beq}{\begin{equation}}
\newcommand{\eeq}{\end{equation}}
\newcommand{\beqa}{\begin{eqnarray}}
\newcommand{\eeqa}{\end{eqnarray}}

\title[Eruptive mass loss]{
Hydrodynamical simulations and similarity relations  for eruptive mass loss from massive stars
}
\author[S.\ Owocki et al.]{
Stanley P.\ Owocki$^1$\thanks{email: owocki@udel.edu},
Ryosuke Hirai$^2$,
Philipp Podsiadlowski$^{2,3}$
\& Fabian R.\ N.\ Schneider$^{4,5,2}$
\\
$^1$
Bartol Research Insitute, 
Department of Physics \& Astronomy, 
University of Delaware, Newark, DE 19716 USA 
\\
$^2$ 
Department of Physics, 
University of Oxford, 
Keble Rd, Oxford, OX1 3RH, United Kingdom
\\
$^3$
Argelander-Institut  f\"ur  Astronomie  der  Universit\"at  Bonn,  
Auf dem  H\"ugel  71,  
D-53121  Bonn,  
Germany
 \\
 $^4$
 Zentrum f\"{u}r Astronomie der Universit\"{a}t Heidelberg,
 Astronomisches Rechen-Institut,
 M\"{o}nchhofstr. 12-14, 
 D-69120 Heidelberg, 
Germany 
 \\
 $^5$
Heidelberger Institut f\"{u}r Theoretische Studien, 
Schloss-Wolfsbrunnenweg 35, 
D-69118 Heidelberg,
 Germany
}

\begin{document}

\include{aas_macros}

\date{Accepted ?.  Received ?; in original form ?}

\maketitle

\label{firstpage}

\begin{abstract}
Motivated by the eruptive mass loss inferred from Luminous Blue Variable (LBV) stars,
we present 1D hydrodynamical simulations of the response from sudden energy injection into the interior of a very massive ($100 \, \Msun$) star.
For a fiducial case with total energy addition set to a factor $f=0.5$ of the net stellar binding energy,
and applied within the stellar envelope, we detail the dynamical response that leads to ejection of the outermost $7.2 \, \Msun$.
We find that the ejecta's variations in time $t$ and radius $r$ for the velocity $v$, density $\rho$,  and temperature $T$ are quite well  fit  by similarity forms in the variable $r/t \approx v$.
Specifically the scaled density follows a simple exponential decline $\rho t^{3} \sim \exp (-r/v_{\rm o} t)$.
This `{\em exponential similarity}' leads to analytic scaling relations for  total ejecta mass  $\DMtot$ and kinetic energy $\DKEtot$ that agree well with the hydrodynamical simulations, with the specific-energy-averaged speed related to the exponential scale speed $v_{\rm o}$ through ${\bar v} \equiv \sqrt{2 \DKEtot/\DMtot} = \sqrt{12} \, v_{\rm o}$,
and a value comparable to the star's surface escape speed, $v_{\rm esc}$.
Models with energy added in the core develop a surface shock breakout that propels an initial, higher-speed ejecta ($>$5000\,km\,s$^{-1}$), but the bulk of the ejected material still follows the same exponential similarity scalings with ${\bar v} \approx v_{\rm esc}$.
A broader parameter study examines how the ejected mass and energy depends on the energy-addition factor $f$,
for three distinct model series that locate the added energy in either the core, envelope, or near-surface.
We conclude by discussing  the relevance of these results for understanding 
LBV outbursts and other eruptive phenomena, such as failed supernovae and pulsational pair instability events.
\end{abstract}

\begin{keywords}
stars: winds, outflows --
stars: mass loss --
stars: massive --
stars: supernovae: general --
shock waves
\end{keywords}

\section{Introduction and Background}

Hot, luminous, massive stars lose mass through strong stellar winds, driven by line-scattering of the star's continuum radiation
\citep{Puls08}.
But for many such massive stars, the overall mass loss may actually be dominated by transient mass ejection episodes associated with eruptive Luminous Blue Variable (eLBV) stars \citep{Smith06b}.
The most prominent and extensively studied eLBV is  the very luminous ($L \approx 5 \times 10^6 \Lsun$), very massive
($M > 100 \Msun$) star $\eta$\,Carinae, which during the ``Great Eruption" of the 1840's is inferred \citep{Smith03b} to have ejected $> 10 \Msun$, resulting in the famous Homunculus nebula observed today.
The associated mass loss rate of $> 1 \Msun$\,yr$^{-1}$ is well above what can be explained by a line-driven wind \citep[$\lesssim 10^{-4.5} \Msun$\,yr$^{-1}$,][]{Smith06b}; but it could in principle be achieved by a quasi-steady {\em continuum}-driven wind associated with the star's super-Eddington luminosity during this epoch \citep{Shaviv00,Owocki04, Owocki17,vanMarle08,Quataert16}.

However, a recent extensive empirical study by \citet{Smith18b} argues that many features of the Great Eruption and the resulting nebula can {\em not} be explained by such a quasi-steady, super-Eddington wind alone, but must also include a relatively sudden {\em explosive} component more analogous to a supernova (SN) explosion than a steady wind.
Indeed, much as inferred for the class of pre-SN eLBV's 
\citep[see review by][and references therein]{Smith14},
the explosive mass loss likely interacts with a more prolonged, quasi-steady, enhanced wind outflow prior to the eruption.
Such pre-SN eruptions may be triggered by core instabilities in the final stages before core collapse  \citep{Chen14}, which lead to energy deposition into the envelope, e.g. by gravity waves \citep{Piro11,Quataert12,Fuller18}, 
or into the core itself, e.g. by pair-instability pulsation \citep{Barkat67,Kasen11,Yoshida16,Takahashi16,Woosley07,Woosley15,Woosley17}, 
or by late-shell-burning instabilities \citep{Smith14b}.
Recent studies by \citet{Gieles18} of the formation of supermassive stars by multiple mergers in the very dense cores of globular clusters also posit a strong mass loss that regulates the maximum mass to be of order 1000\,$\Msun$.

For $\eta$\,Carinae,  the bipolar form of the Homunculus, along with its irregular equatorial ``skirt", has led to broad conjecture 
\citep[][]{Podsiadlowski13,PortegiesZwart16,Smith18b} that the eruption might have stemmed from the {\em merger} of two massive stars
\citep{Gallagher89,Morris06,Morris07,Morris09,MacLeod18}.
In such a merger scenario, the orbital angular momentum leads during a common envelope phase
\citep{Ivanova13,Podsiadlowski13,Justham14} to mechanical ejection of the equatorial skirt, 
while the bipolar Homunculus stems from the {\em energy} deposited into the envelope from the merging of the stellar cores.
This energy addition leads to a prompt ejection of mass, followed perhaps by a strong quasi-steady wind  driven by the super-Eddington luminosity of the post-ejection star.
While both explosive SNe and steady stellar winds have been quite extensively studied through detailed dynamical models, there have been relatively limited attempts to study the dynamics of such sudden mass eruptions
\citep[see, however,][]{Dessart10,Ro13,Ro17}.

A full simulation of such a stellar merger should be 3D, to account for the inspiralling of the two stars, the associated spin-up of the common envelope, and eventual merging of the stellar cores.
This could naturally lead to a distinctly non-spherical form of the ejected nebula, including perhaps both a thin equatorial ejecta along with an associated pinching of the mass ejection into a bipolar form, much as seen in the skirt plus Homunculus of $\eta$\,Carinae.
But to build insight for such complex and computationally expensive 3D models, we elect here to focus first on just the effect of the energy addition associated with the merger of the stellar cores, within the context of idealized 1D, spherically symmetric models.

Specifically, the paper here carries
out a series of 1D time-dependent hydrodynamical simulations of the 
disruption that results when a generic energy addition of unspecified origin is introduced impulsively into a massive 
star and core.
A key feature is that, unlike supernova explosions, the level of energy addition here is taken to be only an order unity factor $f$ of the stellar binding energy $\Ebind$,  thus leading to only partial disruption of the stellar envelope. While some mass fraction is ejected to full escape from the star,   the bulk of the star's mass remains bound by the stellar gravity, and so falls back onto a somewhat inflated surviving star \citep[somewhat analogous to what is thought to occur in {\em failed} explosions from core collapse, e.g.][]{Lovegrove13,Coughlin18b}.
This is intended as just a first step toward future models grounded in specific mechanisms for energy addition, and in the case of merger models, including 2D and 3D dynamics that account explicitly for the role of angular momentum transport in mass ejection and envelope spin-up.

Section 2 describes the basic method and setup for our fiducial model of a $\Mstar = 100 \Msun$ star with impulsive addition of a fraction $f=1/2$ of its binding energy, deposited over the stellar envelope above the core.
The key results in section 3 detail how this unbinds the outermost $\sim 7.2\%$ of the star's mass into an expanding ejecta, with the rest remaining gravitationally bound and thus falling back onto an inflated stellar envelope.
The analysis in section 4 then shows that the long-term evolution of this ejected mass can be well described by relatively simple, analytic similarity forms for the velocity, density and temperature.
Section 5 explores the effect of adding energy {\em within} (vs.\ above) the core, for cases with both half and double the stellar binding energy.
Section 6 extends this to a broad
parameter study for how the mass and kinetic energy of the ejecta scales with energy addition factor $f$ for core, envelope and near-surface locations of the energy addition.
We conclude (section 7) with a discussion of the implications of our results for understanding observed LBV's, and an outline for future work.

\section{1D  Simulation of Envelope Ejection from Sudden Energy Addition}
\subsection{Conservation equations}

The simulations here use a  one-dimensional (1D), spherically symmetric implementation of the multi-dimensional Eulerian hydrodynamics code described by \citet[][]{Hirai16}.
In its full multi-D form a key distinguishing feature of this code lies in its fast and accurate gravity solver; but in the 1D form used here, the gravity  is trivially solved in spherical symmetry.
The code  is otherwise quite similar to other hydro codes  \citep[e.g. {\tt PLUTO},][]{Mignone07} that use Godunov-type schemes to integrate finite-volume forms of the spherical equations for conservation of mass, momentum and energy,
and thereby follow the time ($t$) and radius ($r$) evolution of mass density $\rho$, radial velocity $v$, and temperature $T$.
Conservation of mass takes the form,
\beq
 \frac{\partial \rho}{\partial t} + \frac{1}{r^2} \frac{\partial (\rho v r^2)}{\partial r} = 0
\, ,
\label{eq:masscon}
\eeq
while momentum conservation yields an equation of motion of the form,
\beq
\frac{Dv}{Dt}  = \frac{\partial v}{\partial t} + v \frac{\partial v}{\partial r} = - \frac{1}{\rho} \frac{\partial P}{\partial r} - \frac{G M_r}{r^2}
\, ,
\label{eq:momcon}
\eeq
with $M_r$ the mass contained within radius $r$, and $G$ the gravitation constant.

Here $\partial/\partial t$ and $D/Dt \equiv \partial/\partial t + {\bf v}\cdot \nabla$ are respectively the (Eulerian) partial derivative at a fixed radius, and the total (Lagrangian) time variation in a frame co-moving with the fluid.
In our 1D simulations, the latter represents the time variation for a fixed mass coordinate $M_r$.
The total pressure 
includes components from both radiation and  gas,
\beq
P=  P_{\rm rad} + P_{\rm gas} =   \frac{a_{\rm rad} T^4}{3} + \frac{\rho kT}{\mu} 
\, ,
\label{eq:Ptot}
\eeq 
with $k$ Boltzmann's constant, $\mu$  the mean molecular weight (in g),
and $a_{\rm rad}$ the radiation constant.
The mean molecular weight distribution over mass coordinate is fixed throughout the simulation, with no mixing assumed.

For energy conservation, we evolve the adiabatic form
\beq
\frac{\partial (\rho v^2/2 + E)}{\partial t} 
+ \frac{1}{r^2} \frac{\partial [(\rho v^2/2 + E + P)  v r^2 ]}{\partial r}
= - \rho v \frac{G M_r} {r^2}
\, ,
\label{eq:encon}
\eeq
where the total internal energy again includes contributions from both radiation and a monatomic ideal gas\footnote{For a monatomic gas the internal energy density is $\frac{1}{2}P_{\rm gas}$ for each of the $n=3$ degrees of freedom, which also sets the adiabatic index $\gamma = (n+2)/n = 5/3$.},  $E =  3 P_{\rm rad} + \frac{3}{2} P_{\rm gas} $.

This adiabatic form ignores radiative diffusion and transport, since in the very optically thick flows that develop here this is much smaller than competing terms from energy advection.
Alternatively, we can use the mass and momentum eqns. (\ref{eq:masscon}) and (\ref{eq:momcon}) to recast this energy conservation (\ref{eq:encon}) into a form for just the {\em internal} energy $E$,
\beq
\frac{DE}{Dt}  
= - \left (E + P \right ) \nabla \cdot {\bf v}
= - \frac{E+P}{r^2} \, \frac{\partial (vr^2 )}{\partial r}
\, .
\label{eq:Econ}
\eeq
This form proves useful for interpreting the computed evolution of internal energy and temperature 
(see Appendix A).

\begin{table}
\caption{Input parameters and results for fiducial model of energy addition to stellar envelope.}
\centering
\begin{tabular}{|| l l l ||}
\hline 
{\em Input parameters:} &&  \\
initial mass & $\Mstar$ & 100 $\Msun $  \\
initial radius & $\Rstar$ & 35 $\Rsun$  \\
surface escape speed & $v_{\rm esc} $ & 1044\,km\,s$^{-1}$ \\
gravitational binding  energy & $E_{\rm g}$ & $-3.31 \times 10^{51}$\,erg\\
internal energy & $E_{\rm i}$ & $+2.65 \times 10^{51}$\,erg\\
nets initial binding energy & $\Ebind= E_{\rm g}+E_{\rm i}$ & $-6.65 \times 10^{50}$\,erg \\
energy addition factor & $f$& $ 0.5$ \\
added energy & $-f \Ebind$ & $+ 3.32 \times 10^{50}$\,erg \\
radius range for added energy& $r$ & 7-35\,$\Rsun$ \\
mass range for added energy& $M_r$ &$> 76 \Msun$ \\
\\
{\em Results:} &&  \\
ejecta mass & $\DMtot$ & 7.2 $\Msun$ \\
ejecta kinetic energy & $\DKEtot$ & $0.93 \times 10^{50}$\,erg \\
energy-averaged speed & ${\bar v} $ & 1125\,km\,s$^{-1}$ \\
similarity speed & $v_{\rm o} $ & 325\,km\,s$^{-1}$ \\ 
similarity density  & $\rho_{\rm o} t_{\rm o}^3$ & $1.65 \times 10^{10}$ g\,cm$^{-3}$\,s$^{3}$  \\
similarity temperature & $T_{\rm o} t_{\rm o}^{1.08}$ & $2.6 \times 10^{11}$ K\,s$^{1.08}$ \\
[0.5ex]
\hline
\end{tabular}
\label{table:1}
\end{table}

\subsection{Fiducial model and numerical specifications} 
To explore the nature of mass ejection in eLBV stars like $\eta$\,Carinae, let us first focus on a fiducial massive-star model with the parameters given in Table \ref{table:1}.
This starts with a $\Mstar = 100 \Msun$ stellar model evolved with the public stellar evolution code MESA 
\citep[v10398;][]{Paxton11,Paxton13,Paxton15,Paxton18} up to the time that  the central hydrogen mass fraction has dropped to $X=0.2$.
The resulting He-enriched core contains about 76\% of the star's mass, but is only about 20\% of its $\Rstar \approx 35 \Rsun$ surface radius.
Because of the prominence of radiation pressure, the star's total internal energy, $E_{\rm i} = 2.65 \times 10^{51}$\,erg, is a substantial fraction (more than the 50\% for virial equilibrium of a pure ideal gas) of the gravitational binding energy, $- E_{\rm g} = 3.31 \times 10^{51}$\,erg; 
the  associated net binding energy of the whole star is thus $\Ebind=- 6.65 \times 10^{50}$\,erg. (See Table 1.)

sOur approach idealizes the complex merger processes by assuming that the structure of the merger product is somewhat similar to that of this fiducial model, with a combined mass similar to $\eta$\,Carinae.
The central hydrogen fraction corresponds to a late case A merger, but we do not wish here to specify the details of the pre-merger binary system, or of the merger process itself. 
Instead, within this assumed 1D structure, we focus only on one key aspect of a stellar merger, namely the energy addition to the merged envelope.
The actual amount and distribution of the additional energy will depend on the details of the merger process.
We do not aim here to reproduce this, but rather
explore the overall effects of varying the location of energy deposition from the core, to the envelope, to the near-surface.

\begin{figure*}
\begin{center}
\includegraphics[scale=.38]{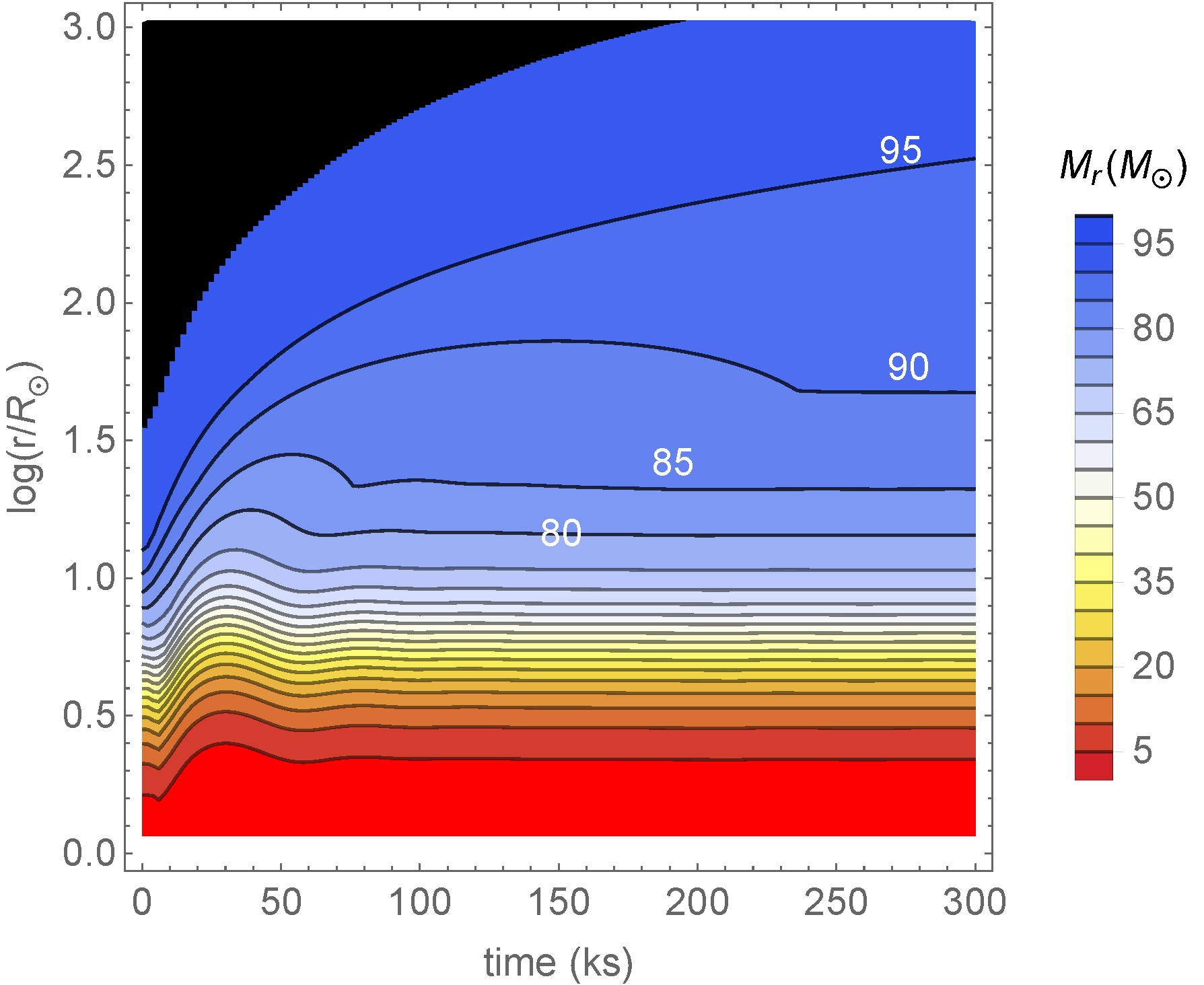}
\includegraphics[scale=.55]{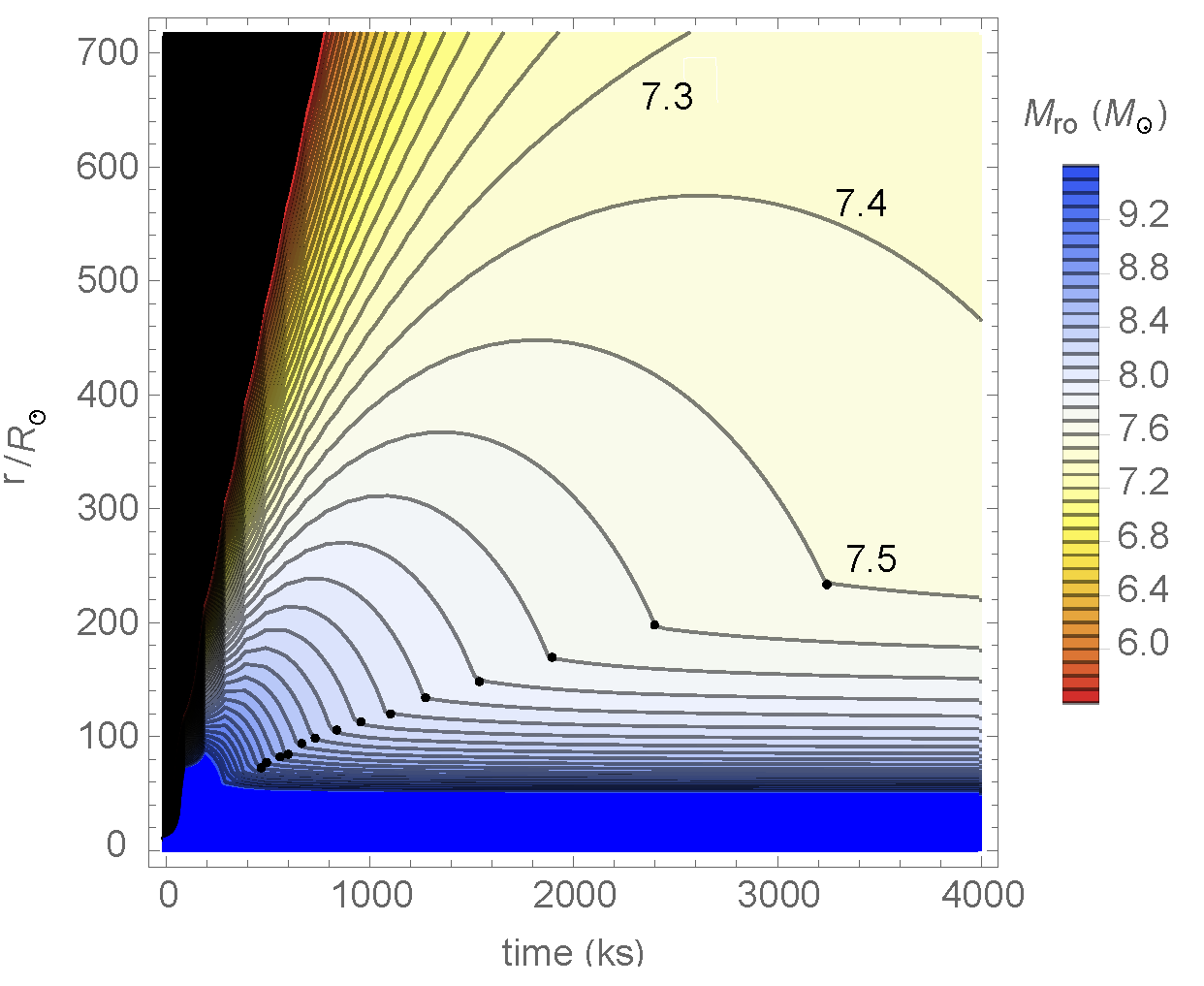}
\caption{{\em Left:} Contours of mass coordinate $M_r$ vs.\  log radius $r$ and time showing early response $t=0-300$\,ks response to impulsive energy addition $\Delta E = -f \Ebind$ applied over the envelope range $r=7-35 \Rsun$ of the initial star.
{\em Right:} Contours of mass to outer boundary $M_{\rm ro} \equiv \Mstar - M_r$, plotted now for linear radius range $r=0-700 \Rsun$ and for the full $t=0-4050$\,ks of the simulation.
The dots indicate the location of re-accretion shocks, where mass (with $M_{\rm ro} > 7.4 $) that was launched upward with insufficient energy to escape has fallen back onto the remnant star.
}
\label{fig:fig1}
\end{center}
\end{figure*}

 The energy directly available from a merger is given by the sum of the binding energies of the two stars, plus their orbital energy,  minus the binding energy of the assumed 100$\Msun$ star.
 Neglecting for simplicity any net change in binding energy as relatively small, the orbital kinetic energy scales with the product of masses divided by the final separation $a$ at the onset of the dynamical stage,  $E_{\rm kin} \approx G M_1 M_2 /2a$.
 For the optimal case of two equal-mass stars with $M_1=M_2=50 M_\odot$, and taking a final separation in the range $a= (7-35) R_\odot$ that corresponds to the core to surface radius of our merged star,  this orbital energy deposition lies in the range $E_{\rm kin}  \approx (1.3-6.5 )\times 10^{50}$\,erg,
 which is an order-unity factor $f$ of the net stellar binding energy of our assumed model.
 In principle, even more energy could arise from the difference in the binding energies, or even from sudden hydrogen burning if original envelope material is mixed into the hot, merged core \citep{Podsiadlowski10}.

 In our simulations we first map the stellar model on the centre of a spherical grid, with
 a dilute atmosphere of negligible dynamical influence placed around the star. 
At an initial time we impulsively add some fraction $f$ of the stellar binding energy to the stellar envelope above this core, distributed in proportion to the mass.
The  analysis below focuses on detailed results for a fiducial $f=0.5$ model, i.e. with an added energy $\Delta  E \approx +3.3 \times 10^{50}$\,erg;
 but, as discussed in section 6,  we have also computed a full grid of models from $f=0.1$ to $2.0$ in steps of $0.1$, with three model series with energy deposition in either the core, envelope, or near surface.
As noted in the introduction, while motivated by the specific notion of a merger for the giant eruption in $\eta$\,Carinae, this choice of a broad range for the energy fraction and location mean that results could also be relevant for eruptive mass loss  from a variety of specific mechanisms.


To follow the evolution of mass ejected well beyond the initial  stellar radius $\Rstar \approx 35 \Rsun$, our numerical grid extends from the stellar center $r=0$ to a  large maximum radius $r=R_{\rm max} = 4000 \Rsun \gg \Rstar $, 
using $nr=25,000$ radial cells with spacing that increases with grid index $i$ as $\Delta r_{i+1} = 1.000186 \, \Delta r_i$, with $\Delta r_0= 0.0072 \, \Rsun$.
We apply outgoing boundary conditions at the outer boundary radius $R_{\rm max} = 4000 \Rsun$.
In order to follow mass outflow to this large radius, our fiducial model is evolved to a very long time, $t_{\rm max} = 4050$\,ks, with time steps set to a maximum of $0.9$ times the Courant time
 \citep{Courant67} .

\section{Simulation Results for Fiducial Model}

\subsection{Mass ejection vs.\ fallback}

Figures \ref{fig:fig1} illustrate the radial evolution of mass shells over the initial $300$\,ks.
The impulsive introduction of energy at initial time $t=0$ is applied over the radial range ($r=7-35 \, \Rsun$), distributed in proportion to the mass.  
The mass contours show that the initial response of nearly all the mass -- even in the core with $M_r <76 \, \Msun$ -- is to expand in radius.  
For most of the mass, this expansion is followed by a partial fallback to a final radius that remains somewhat above the initial radius.

The outer $ \sim 7.2 \Msun$ of the envelope escapes completely from the star.
The right panel plots contours of mass measured from $r$ outward, i.e. $M_{\rm ro} \equiv \Mstar -M_r$, 
using now a linear radius scale from $r=0-700 \Rsun$, and time plotted over the full $t=0-4$\,Ms of the simulation.
This shows that a parcel with $ M_{\rm ro} =7.5 \Msun$ rises to  $r>500 \Rsun$ before falling back;
indeed, we shall see that even the still-rising parcels with $M_{\rm ro} = 7.3$ and 7.4 $\Msun$ have a slightly negative net total energy, and so will eventually fall back. But parcels with $M_{\rm ro} \lesssim 7.2 \Msun$ 
($7.2$\% of the stellar mass) do completely escape.

\begin{figure*} 
\begin{center}
\includegraphics[scale=.36]{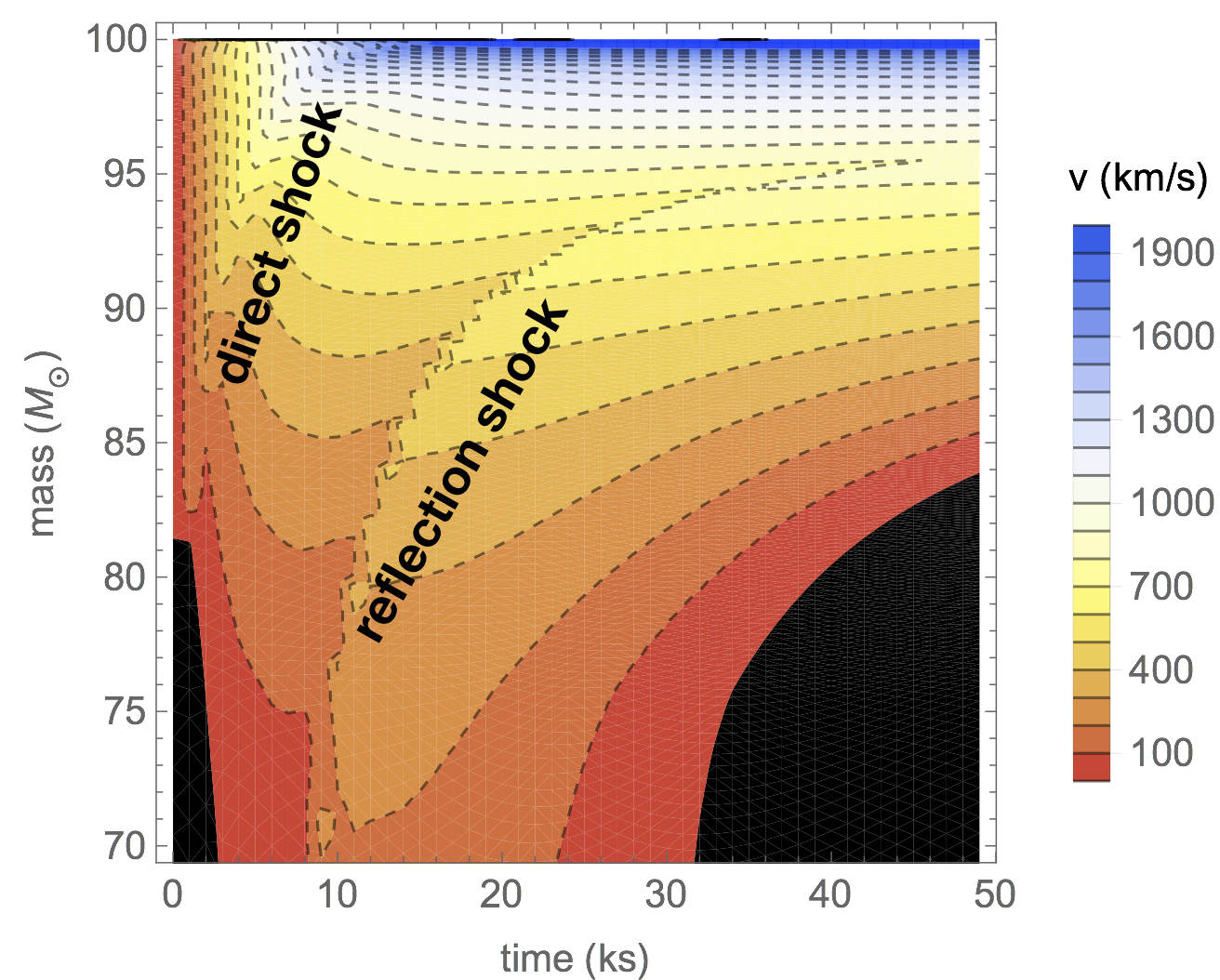}
~~~
\includegraphics[scale=.32]{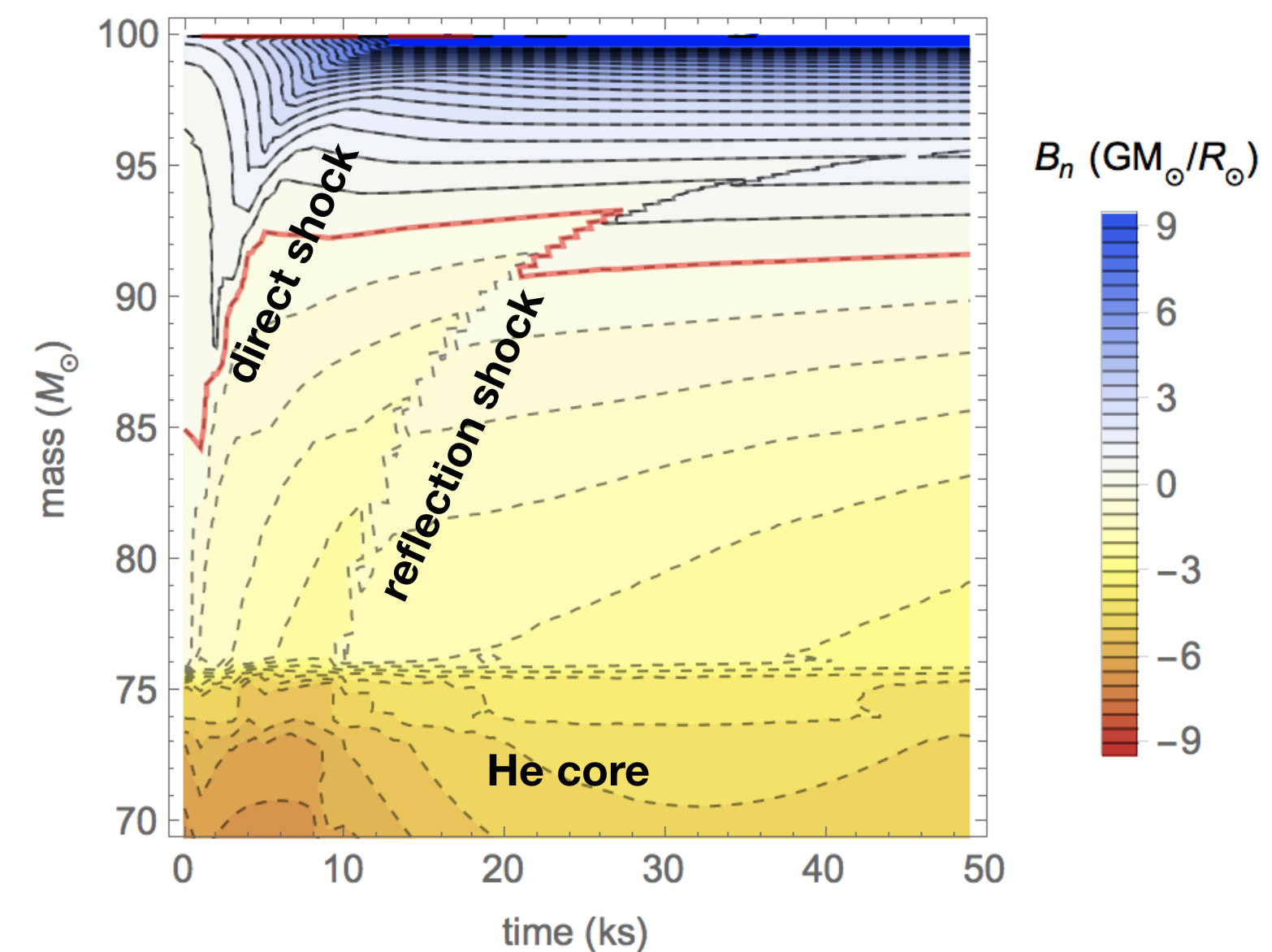}
\caption{{\em Left:} Velocity evolution plotted as contours over mass  range $M_r = 70-100 \Msun$ and initial $t=50$\,ks.
{\em Right:} Contours of associated Bernoulli energy $B_{\rm n}$ over the same mass and time ranges,
with solid and dashed contours for respectively positive and negative $B_{\rm n}$, and zero contour highlighted in red.
The annotations mark the direct and reflection shock propagation, and the
He core below the envelope heating region.
}
\label{fig:fig2}
\end{center}
\vskip -0.2in
\end{figure*}

\subsection{Variation of Bernoulli energy with mass and time}

To explore further how this mass ejection is achieved, the left panel of figure \ref{fig:fig2} shows a contour plot of velocity $v$ vs. mass $M_r$ 
over the early time interval $t=0-50$\,ks.
As denoted by the colorbar, the colors from red to blue indicate positive (outflowing) velocities in the range $v= 0-1900$\,km/s;
 the black in the lower right corner denotes mass and time where $v<0$, i.e. infall.

These velocity contours clearly show two propagating discontinuities, which we identify with  direct and rebound shocks that arise from the sudden initial deposition of energy in the mass range $M_r =76-100 \, \Msun$.

The right panel of figure \ref{fig:fig2} shows a corresponding contour plot of the total specific energy, also known as the  Bernoulli energy,
\beq
B_{\rm n} \equiv \frac{v^2}{2} - \Phi + h
\, ,
\label{eq:Bndef} 
 \eeq
where the gravitational potential is defined by
\beq
\Phi \equiv \frac{G M_{r}}{r} 
+ 4 \pi G \int_r^\infty \, r' \rho(r')  \, \textrm{d}r'
\, ,
\label{eq:phidef}
\eeq
and 
\beq
h \equiv \frac{E+P}{\rho} =  \frac{5}{2}  \frac{kT}{\mu} +  \frac{4a_{\rm rad} T^4}{3 \rho}
\, 
\label{eq:hdef}
\eeq
 is the total specific enthalpy due to both gas and radiation.
This is defined in terms of the associated total internal energy $E$ and total pressure $P$, and
given in the second equality in terms of density $\rho$, temperature $T$, and associated constants.
The colorbar shows the numerical value of $B_{\rm n}$ (in units of $G \Msun/\Rsun$), 
with white to blue being positive, and yellow to red being negative.

The red boldface contour more explicitly marks the divide between the upper-right region in time and mass, where this total energy is positive ($B_{\rm n}>0$), vs. the lower regions where it is negative ($B_{\rm n}<0$).  
This roughly separates mass parcels that will escape the star from those that will fall back to it.

\begin{figure}
\begin{center}
\includegraphics[scale=.55]{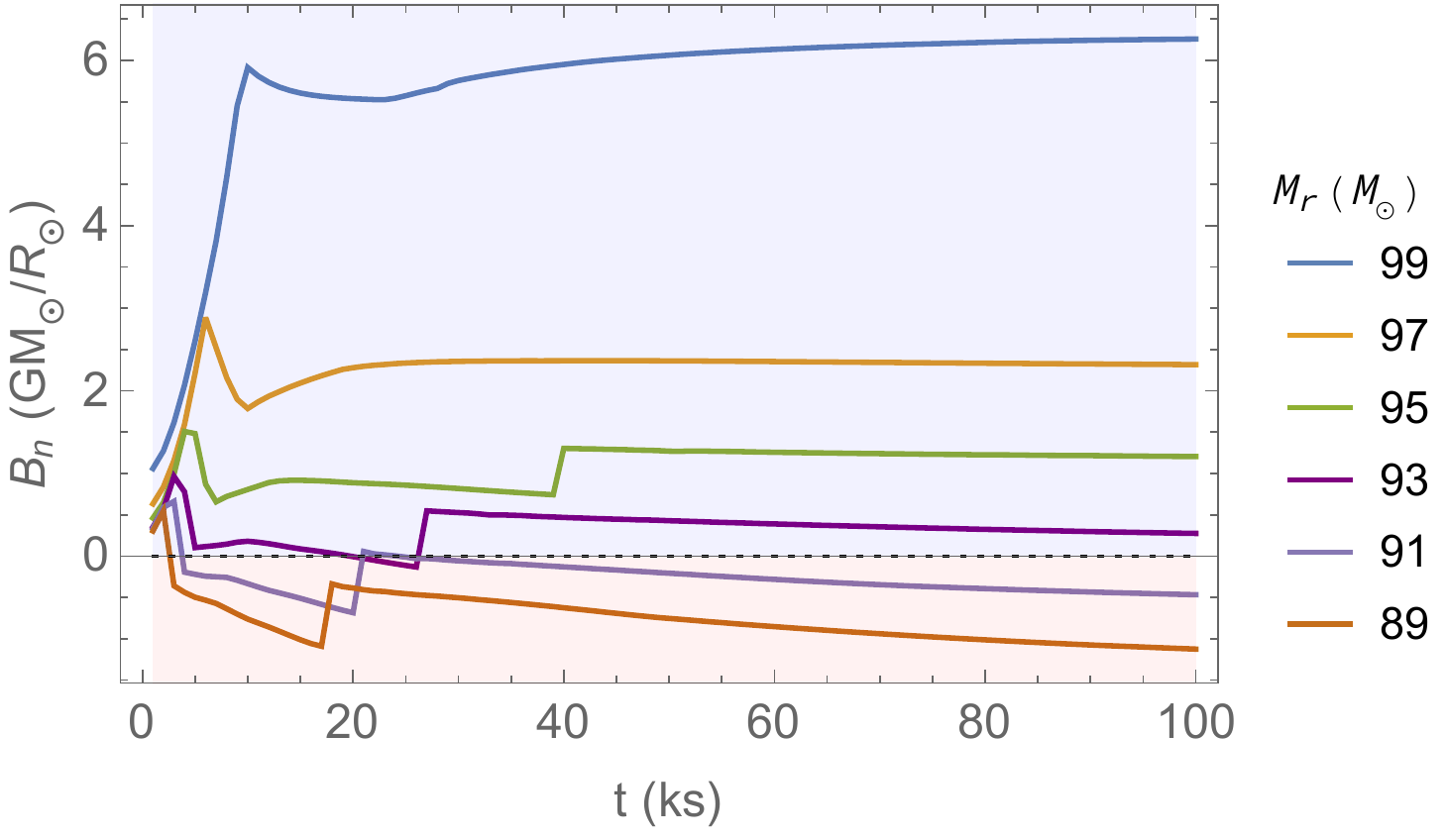}
\caption{For mass parcels with fixed $M_r$ given in the right legend, the time evolution of Bernoulli energy $B_{\rm n}$, with colors highlighting the division between bound ($B_{\rm n} < 0$; red region) vs.\ unbound ($B_{\rm n} > 0$; blue region) parcels.
}
\label{fig:fig3}
\end{center}
\end{figure}

\begin{figure}
\begin{center}
\includegraphics[scale=.56]{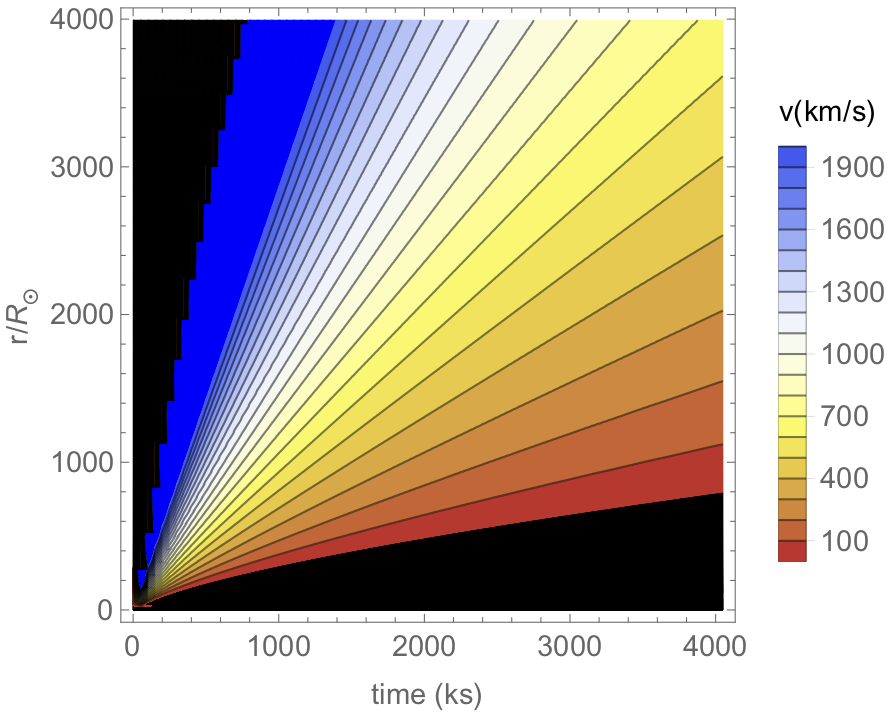}
\includegraphics[scale=.5]{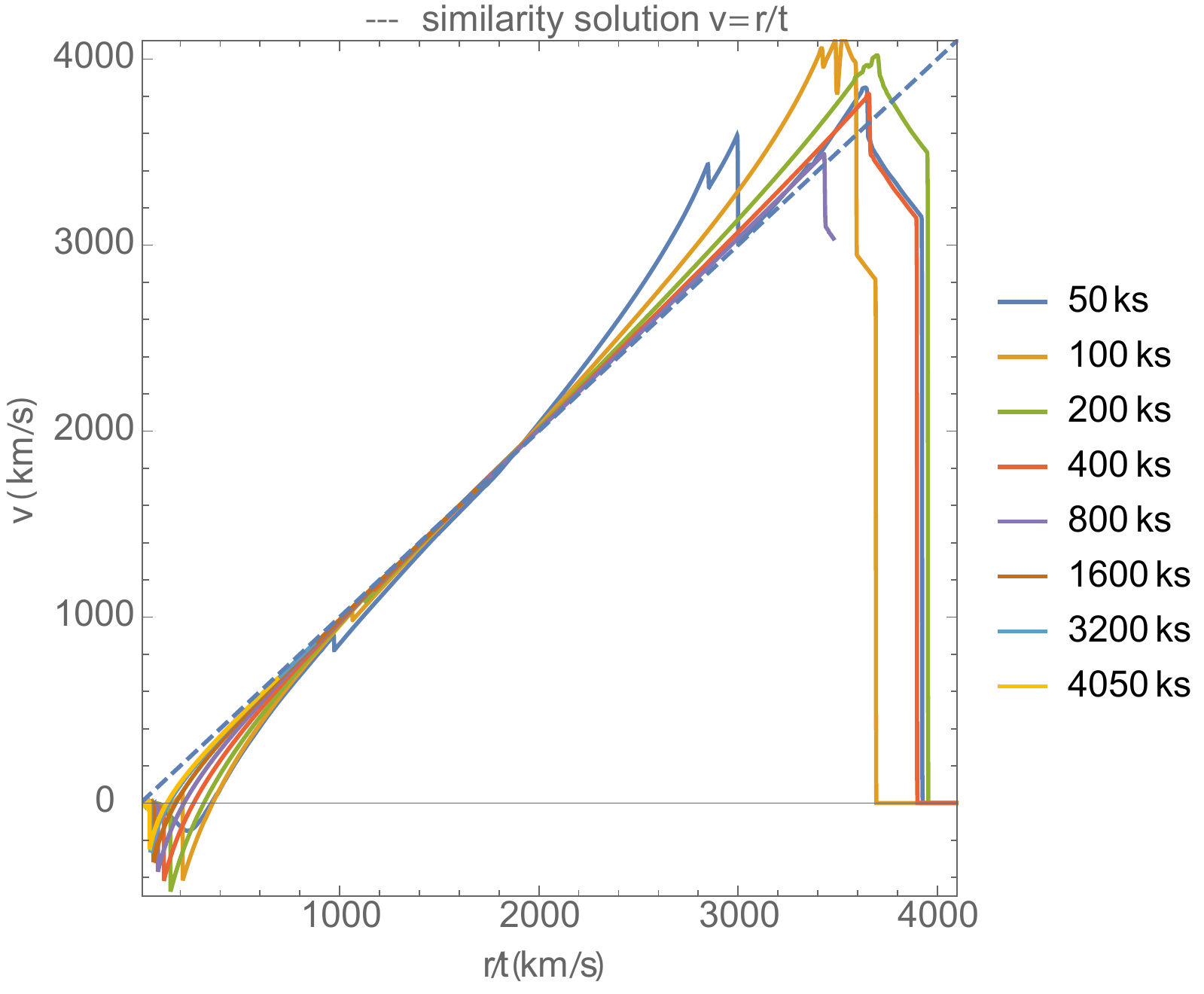}
\caption{{\em Top:} Contours of velocity  over the full range in radius $r=0-4000 \Rsun$ and time $t=0-4050$\,ks of the simulation, showing a  fan-like expansion indicative of a similarity form $v = r/t$.
{\em Bottom:} Plots of $v$ vs. $r/t$ for the indicated time snapshots. At early times, the initial ejecta forms shocks in the ambient medium, eventually clearing it beyond the outer boundary. At late times, the ejecta velocities asymptotically approach the similarity form, $v = r/t$, shown by the  dashed line. 
The negative velocities for small $r/t$ reflect the fall-back of failed mass ejecta.}
\label{fig:fig4}
\end{center}
\end{figure}

For a  set of mass parcels with mass coordinate  in the range $M_r = 89-99 \Msun$ (as marked in the right legend),
figure \ref{fig:fig3} plots the corresponding Lagrangian variations in time $t$ of this Bernoulli energy $B_{\rm n} (t,M_r)$.
Note that following the boosts from the two shocks, $B_{\rm n}$ for each parcel remains nearly constant, 
but  with just a slow decline for bound parcels.

The reason for this decline can be understood through analysis of the time-dependent equations for conservation of mass and energy, 
eqns. (\ref{eq:masscon}) and (\ref{eq:encon}).
With some manipulation, these can be recast to show that
\beq
\frac{D B_{\rm n}}{Dt} = \frac{1}{\rho} \frac{\partial P}{\partial t}  - \frac{\partial \Phi}{\partial t}
\, ,
\label{eq:DBnDt}
\eeq
where again $D/Dt$ represents the total Lagrangian variation co-moving with the flow.

In a steady-state flow with all $\partial/\partial t=0$, this means $B_{\rm n} =$\,constant along the flow of a mass parcel.
But  if the right-hand-side of eqn.\, (\ref{eq:DBnDt}) is negative, then $B_{\rm n}$ will decline in time within a mass parcel, as seen in figure \ref{fig:fig3} for the bound parcels.  
At later times, these intrinsic time variations vanish, and so the Bernoulli energy over the full range of our model, extending to the maximum radius $R_{\rm max} = 4050 \Rsun$ and final time $t_{\rm fin} = 4050$\,ks, does indeed approach a fixed constant value for each mass parcel of the ejecta.
Specifically, solving $B_{\rm n} (\DMtot, t_{\rm fin}) \equiv 0$ from our simulation gives $\DMtot = 7.18 \Msun$ as an estimate for the total ejecta mass in this fiducial model.

\begin{figure}
\begin{center}
\includegraphics[scale=.5]{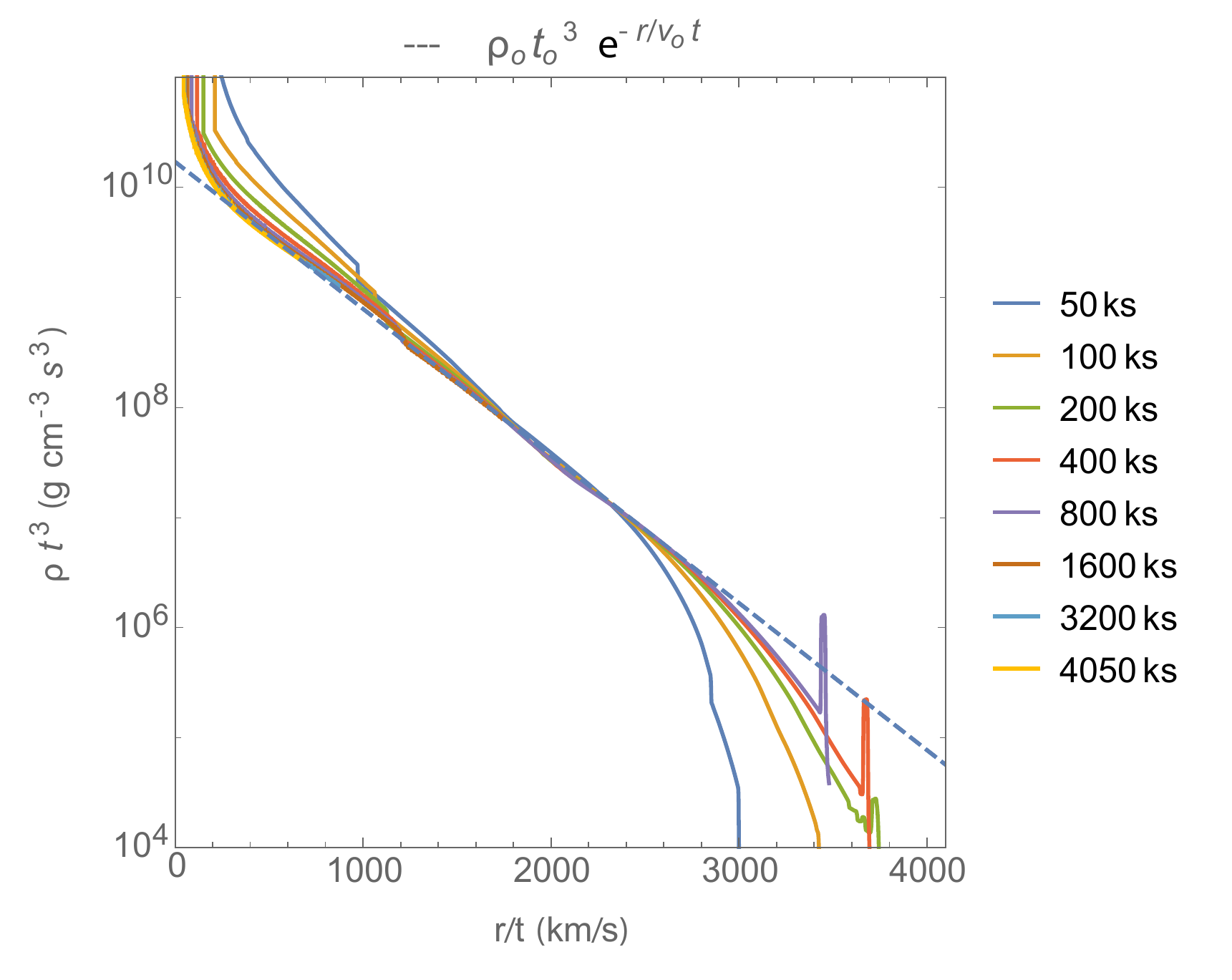}
\caption{Plots of time-scaled density $t^3 \rho$ vs. similarity variable $r/t$ for the indicated time snapshots. 
The dashed line shows the visually best-fit line, representing exponential decline $e^{-r/v_{\rm o} t}$, with $v_{\rm o} \approx 325$\,km\,s$^{-1}$, and density normalization given in Table 1.
}
\label{fig:fig5}
\end{center}
\end{figure}

\section{Similarity relations}
\label{sec:simsolns}

\subsection{Free-expansion velocity of ejecta}
\label{eq:simvel}

Over this full range in radius $r=0-4000 \Rsun$ and time $t=0-4.05$\,Ms, the top panel of figure \ref{fig:fig4} plots contours of the velocity $v$.
The resulting overall fan-like form of these contours suggests that this velocity variation can be approximated by a similarity form $v(r,t) \approx r/t$.
The bottom panel of figure \ref{fig:fig4} shows line plots of $v$ vs.\ $r/t$ for the selected time snapshots denoted in the righthand legend.
Note that indeed, for all but the very early time $t=50$\,ks, the velocity approaches the similarity form $v(r,t) = r/t$ marked by the dashed line.

This similarity form for velocity is just what is found in free-expansion, such as occurs in supernova explosions, or even in the Hubble law for the expanding universe.
However, it is important to emphasize that in this case it applies only approximately for unbound material that escapes from the star, and then only at large $r/t$, and for times $t  \gg 100$\,ks.

\subsection{Exponential similarity relation for density}

Since in such a free expansion, the specific volume increases as $r^3 \sim t^3$, one expects the density to decline as $\rho \sim 1/t^3$.
Figure \ref{fig:fig5} thus shows plots of the time-scaled density $\rho t^3$ vs.\ the similarity variable $r/t$.
Note that this scaled density is not constant, but declines with $r/t$, approaching the dashed line at large $r/t$ and larger times $t$.
On this semi-log scale, such a linear decrease indicates an {\em exponential} decline, represented here by the dashed line for the quoted fit function,
\beq
\rho (r,t) \approx \frac{\rho_{\rm o} t_{\rm o}^3}{t^3}
e^{-r/v_{\rm o} t}
\, ,
\label{eq:rhofit}
\eeq
where we find the visually best fit parameters 
$\rho_{\rm o} t_{\rm o}^3 \approx  1.65 \times 10^{10}$ g\,cm$^{-3}$\,s$^{3}$ and $v_{\rm o} \approx 325 $\,km/s. 
Such an exponential similarity relation for density has also been noted previously for SNIa explosions \citep{Dwarkadas98}, but in the present context of eruptions we argue below (sections 5 and 6) that this may be a signature of the role of gravity in slowing the ejecta and keeping some of the stellar mass bound.s
Appendix A shows that analogous exponential similarity relations hold here for the internal energy and temperature.

\subsection{Mass distribution of ejecta velocity}

We can use the density fitting formula (\ref{eq:rhofit}) to derive an associated estimate for the time and space variation of ejected mass,
\beqa
 \Delta  M (r,t) 
 &=& 4 \pi \int_r^\infty \,r'^2 \rho(r',t)  \, \textrm{d}r'
 \nonumber
 \\
&=& 4 \pi \frac{\rho_{\rm o} t_{\rm o}^3}{t^3} \int_r^\infty \,  r'^2
 e^{-r'/v_{\rm o} t}
 \, \textrm{d}r'
 \nonumber
 \\
&\equiv& \DMtot \, m(r/v_{\rm o} t)
 \, ,
 \label{eq:DMrt}
 \eeqa
where 
the total ejected mass is
 \beq
\DMtot \equiv 8 \pi \rho_{\rm o} t_{\rm o}^3 v_{\rm o}^3
= 7.16 \Msun
 \, ,
 \label{eq:DMtot}
 \eeq
 with the latter evaluation applying the above fitting values for $\rho_{\rm o} t_{\rm o}^3$ and $v_{\rm o}$.
The resulting total mass agrees very well with the net mass ejection $7.2 \Msun$ inferred from the numerical simulation,
 as indicated by figure \ref{fig:fig1}b.
The spatial and temporal distribution of the ejected mass also follows a similarity form, given in terms of the dimensionless similarity variable $x=r/ v_{\rm o} t$ by
 \beq
 m(x) \equiv \left (1 +x + \frac{x^2}{2} \right ) 
 e^{-x}
 = \frac {\Mstar - M_r}{\DMtot}
 \, ,
 \label{eq:mxdef}
 \eeq
 where the latter expression gives the relationship to the standard mass coordinate $M_r$.
This represents a {\em cumulative distribution function} for the fraction of mass ejected above a scaled velocity $x=v/v_{\rm o}$.
 
 If we  next apply the similarity form for the velocity $v=r/t$, we see that $m(v/v_{\rm o})$ also gives the mass dependence on ejecta velocity.
 Defining then the {\em inverse} function $\chi(m) \equiv m^{-1} (\chi)$, we can write a similarity form for the mass distribution of ejecta velocity,
 \beq
 v(M_r) = v_{\rm o} \,  \chi \left (  \frac {\Mstar - M_r}{\DMtot} \right )
 \,  .
 \label{eq:vMr}
\eeq
\begin{figure}
\begin{center}
\includegraphics[scale=.5]{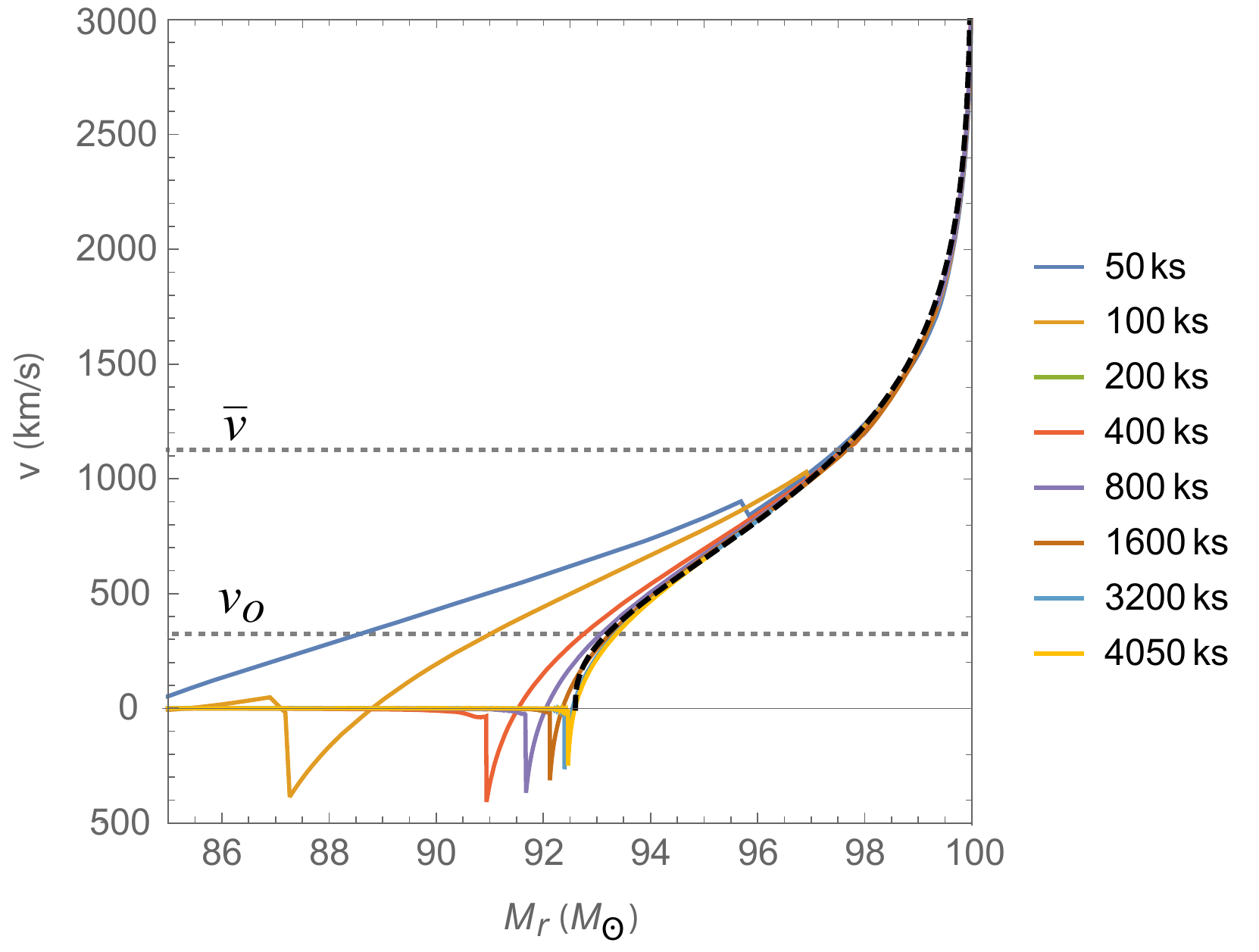}
\caption{Plots of velocity distribution in mass $M_r$ for  the indicated time snapshots.
For later times when marginally bound material  with $M_r \lesssim 92.6 \Msun$ has fallen back to the star, the ejecta speeds approach the analytic similarity form from eqn.\ (\ref{eq:vMr}) (dashed curve).
The lower and upper horizontal dotted lines compare the density cutoff speed $v_{\rm o}$ and the average speed ${\bar v}= \sqrt{12} v_{\rm o}$ defined in eqn. (\ref{eq:vbar}).
The rightmost spike indicates that a small amount of material is ejected at speeds much higher than this mean.
}
\label{fig:fig6}
\end{center}
\end{figure}
Figure \ref{fig:fig6} compares plots for the velocity vs. mass for the indicated time snapshots of the our full hydrodynamical simulation;
the dashed curve represents the associated analytic  form $v(M_r)$ from eqn.\, (\ref{eq:vMr}).
For large times and masses above the marginally unbound mass with $M_{r} = \Mstar- \DMtot = 92.8 \Msun$, there is again remarkably close agreement between the analytic similarity relation and the numerical simulation results.
 Note that only a small amount of mass reaches the highest speeds $v > 3000$\,km\,s$^{-1}$.
 
 \subsection{Total kinetic energy of ejecta}
 
 We can also use these similarity relations to obtain an analytic scaling for the total kinetic energy of the ejected material,
 \beqa
 \DKEtot
 &=& 4 \pi \int_0^\infty \,\rho(r,t) \frac{v^2}{2} r^2 \, \textrm{d}r
 \nonumber
 \\
&=& 2 \pi \frac{\rho_{\rm o} t_{\rm o}^3}{t^5} \int_0^\infty \,  r^4
 e^{-r/v_{\rm o} t}
 \, \textrm{d}r
 \nonumber
 \\
&=& 2 \pi  \rho_{\rm o} t_{\rm o}^3 \, v_{\rm o}^5 \int_0^\infty \,  x^4 e^{-x} \, \textrm{d}x
 \, ,
  \nonumber
 \\
&=& 6 \, \DMtot v_{\rm o}^2 \approx 9.3 \times 10^{49} \, {\rm erg}
 \, ,
 \label{eq:DKEtot}
 \eeqa
 where the final numerical evaluation is again in quite good agreement with what's found in the full numerical simulations.
 It is also comparable to the kinetic energy inferred for the ejecta in $\eta$\,Carina \citep{Smith03}.

The scaling in the last line of (\ref{eq:DKEtot}) shows that the characteristic speed $v_{\rm o}$ from the exponential truncation of density is the key parameter setting the specific kinetic energy of the ejecta, $\DKEtot/\DMtot$.
The associated averaged speed is
\beq
{\bar v} \equiv \sqrt{\frac{2  \, \DKEtot}{\DMtot} } = \sqrt{12} v_{\rm o} = 1125 ~ {\rm km \, s^{-1}} 
\approx v_{\rm esc}
\, .
\label{eq:vbar}
\eeq
We show in  section 6.1 (see figure \ref{fig:fig9}b) that the connection in the last equality between the average ejecta speed and the star's surface escape speed is a quite general property for eruption models with moderate energy factors $f<1$.

\begin{figure*}
\begin{center}
\includegraphics[scale=.36]{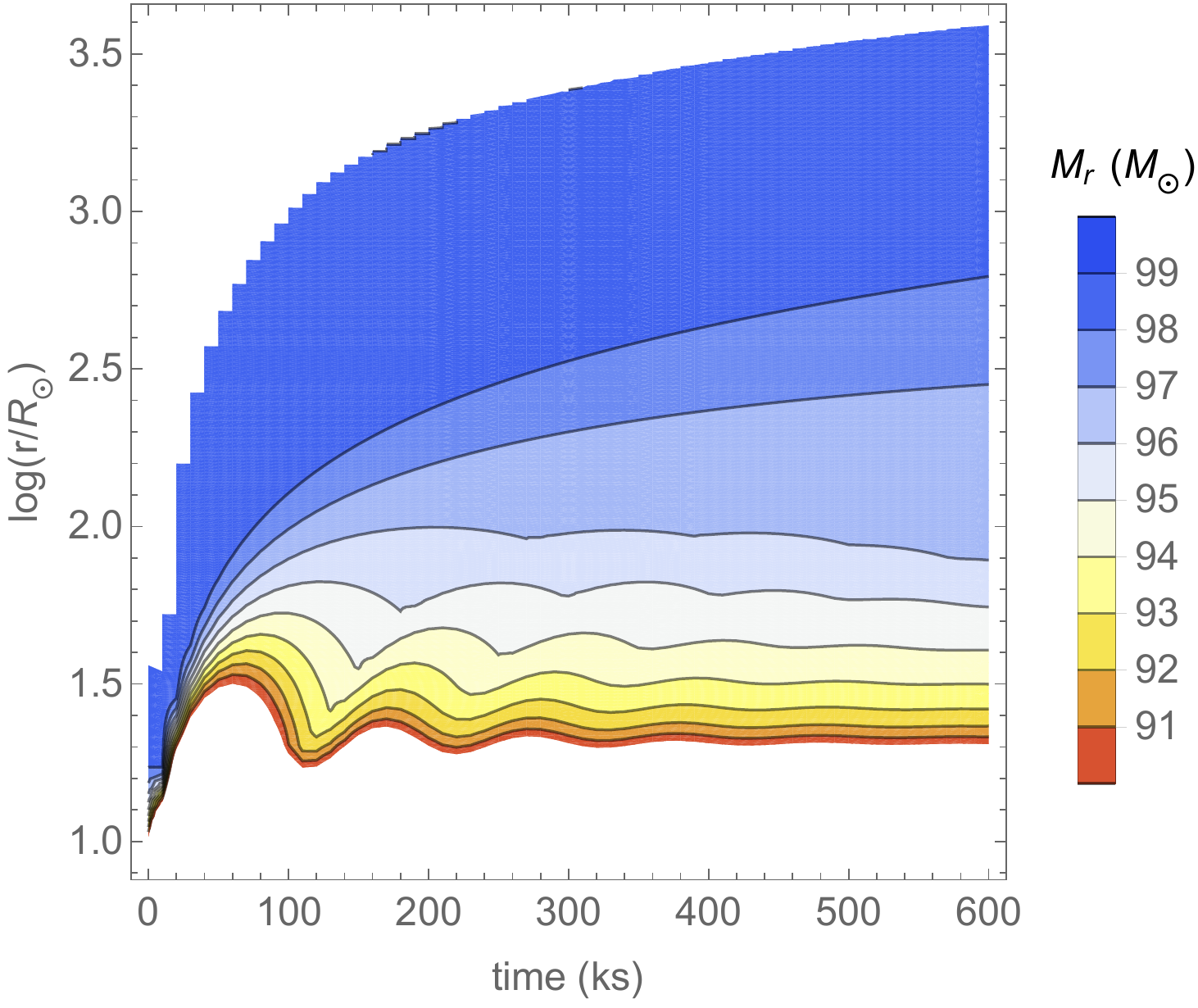} ~~~
\includegraphics[scale=.36]{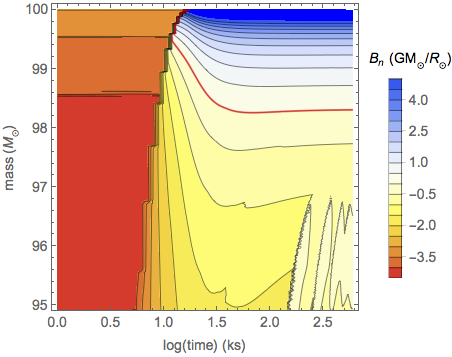}~
\includegraphics[scale=.36]{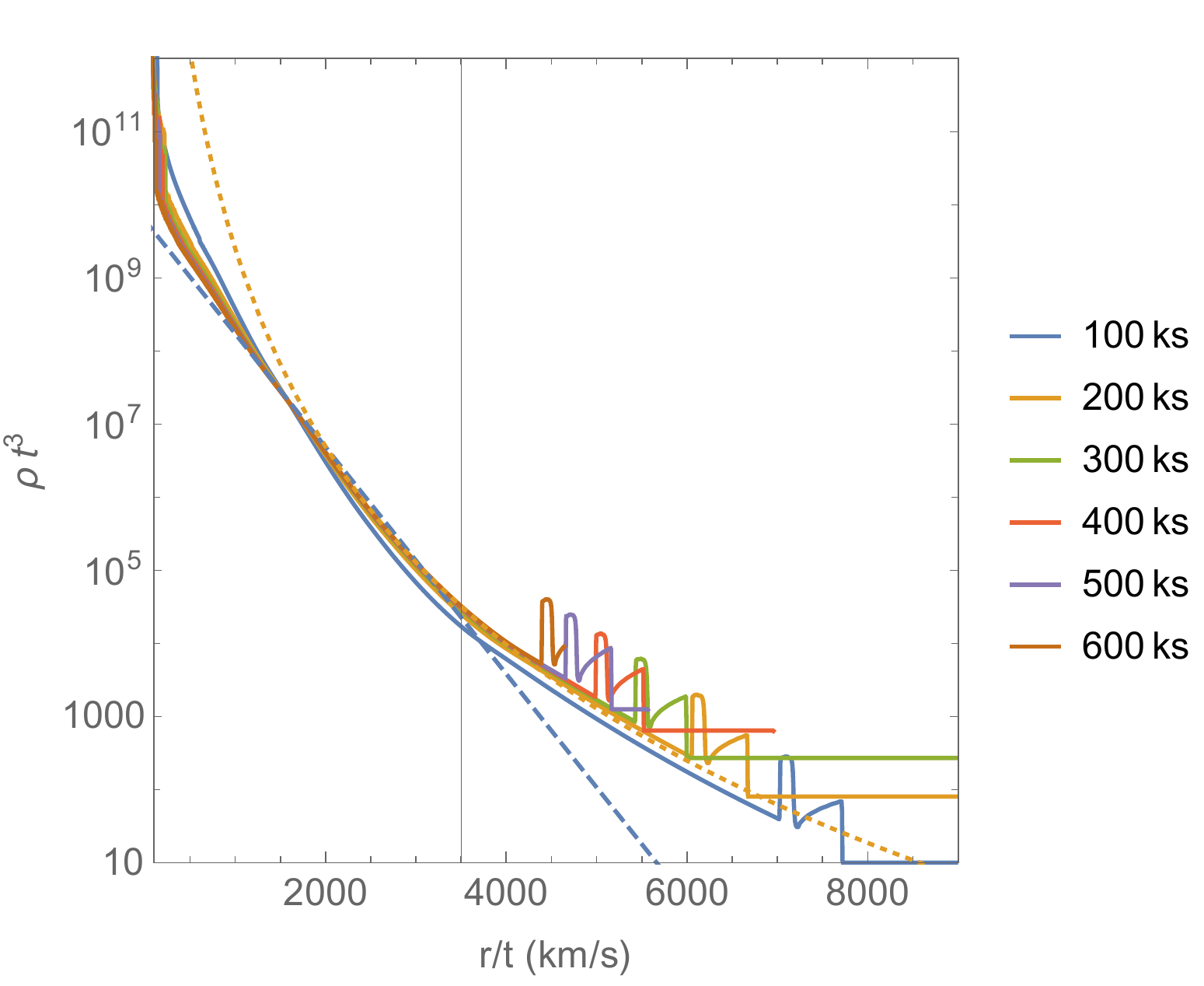}
\caption{
Results for $f=0.5$ model with energy addition in the stellar core ($M_{r} < 76 \Msun$).
{\it Left}: Mass contours vs. time and log radius.
{\it Middle}: Variation of Bernoulli energy vs. mass and log(time).
{\it Right}: Plots of scaled density vs. similarity speed $r/t$ for the cited times. 
The dashed line shows the visual best-fit for the exponential decline at small speed, while the dotted  curve shows a power-law fit to the high-speed tail above a `breakout' speed $v_{\rm p} = 3500$\,km\,s$^{-1}$ (vertical line).
}
\label{fig:fig7}
\end{center}
%
\begin{center}
\includegraphics[scale=.36]{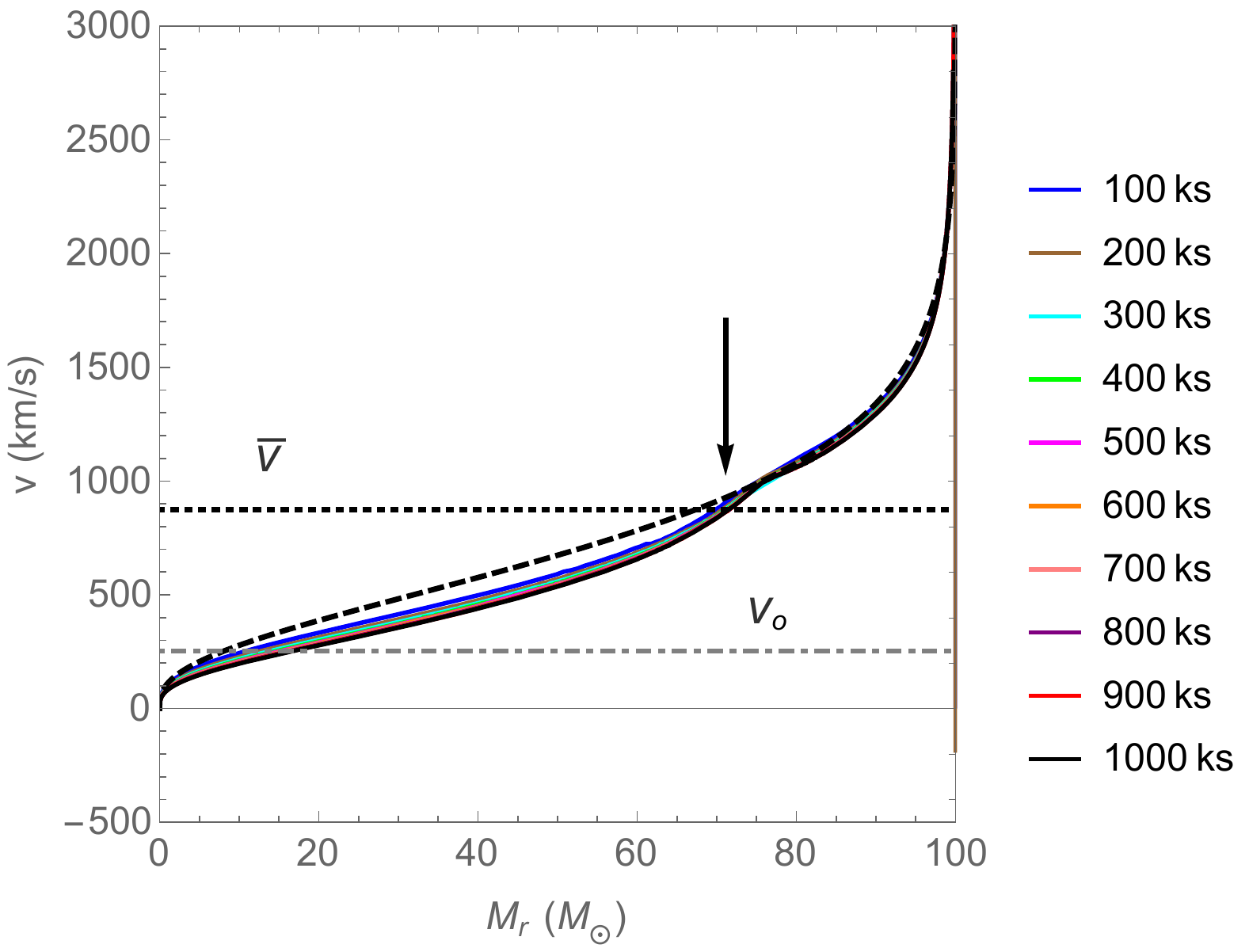}
\includegraphics[scale=.18]{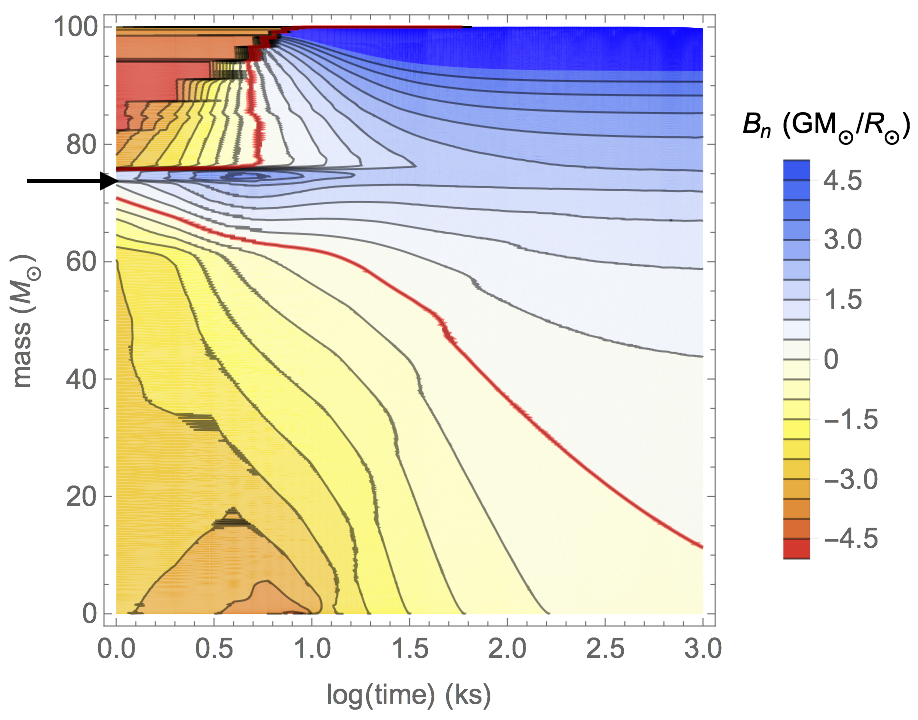}
\includegraphics[scale=.36]{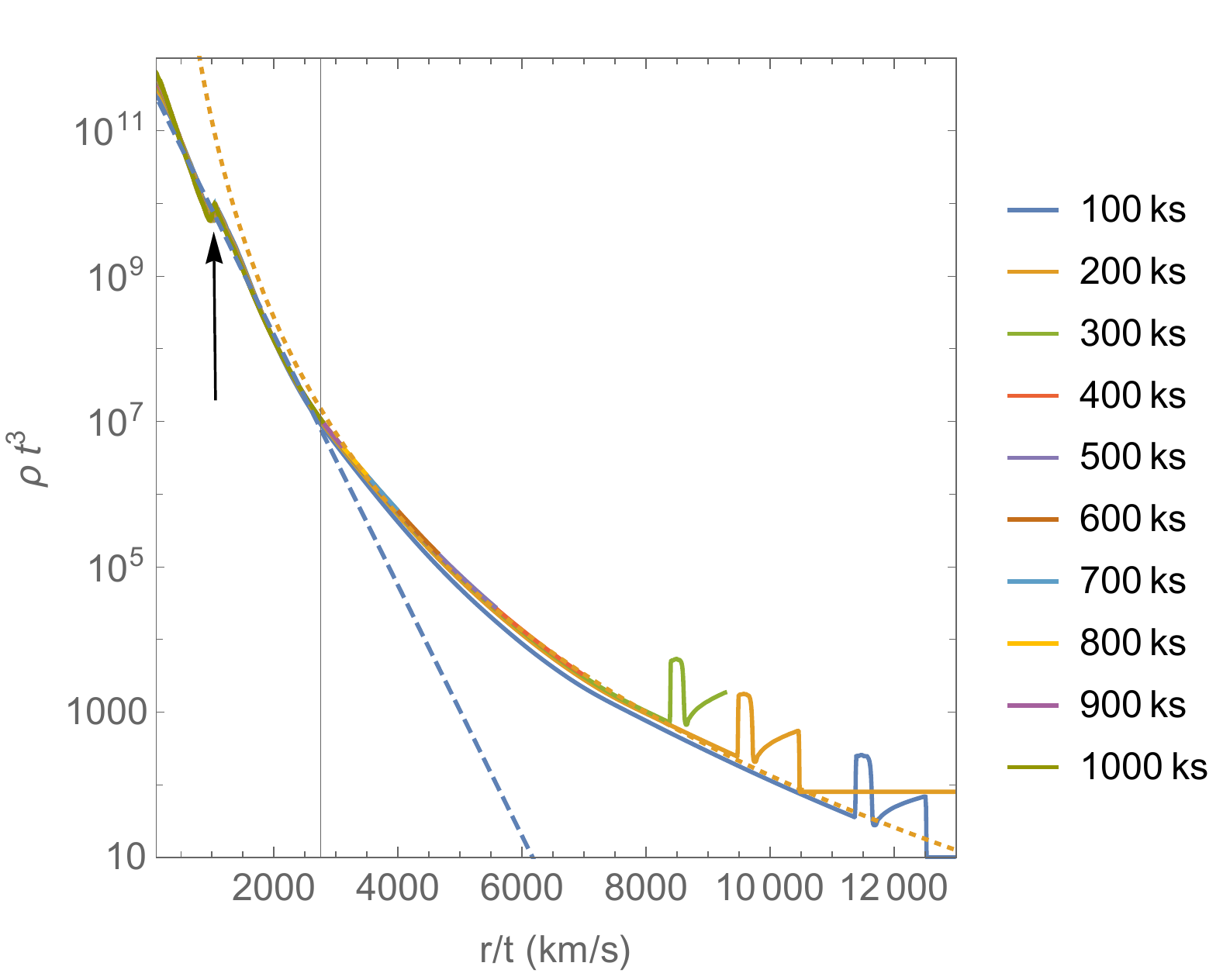}
\caption{Results for core-energy addition with factor $f=2.0$, i.e., {\em twice} the stellar binding energy.
As in figure \ref{fig:fig7}, the middle and 
right
panels show respectively the mass and time evolution of the Bernoulli energy, and the variation of scaled density with similarity speed $r/t$;
but  now the initial ejecta extends to very high $B_{\rm n}$ (blue color at top of middle panel) with associated very high speeds, fit by power-law (dotted curve in the bottom panel) that now extends well beyond the exponential fit (dashed curve) at lower speeds.
The 
left
panel now compares the mass distribution of velocity for the cited time snapshots, along with the analytic similarity form (dashed curve).
The dotted and dot-dashed horizontal lines mark the mean speed ${\bar v} =873$\,km\,s$^{-1}$ and the exponential similarity speed $v_{\rm o} = {\bar v}/\sqrt{12} = 252$\,km\,s$^{-1}$.
In all three panels the arrows mark the location of the upper boundary of the heated core, with $M_{r} \approx 76 \Msun$.}
\label{fig:fig8}
\end{center}
\end{figure*}

Since the ejecta is gravitationally unbound, and generally also highly supersonic,  the kinetic energy essentially represents the {\em total} energy of the ejecta. 
From total energy conservation between the initial and final state, we thus see that the final binding energy of the remaining star just depends on the kinetic energy and the net initial energy,
 \beq
\BEfin \approx (1-f) \Ebind - \DKEtot = -  4.2 \times 10^{50} \, {\rm erg}
 \, .
 \label{eq:BEfin}
 \eeq 
 As this is less negative than the binding energy $\Ebind = -6.65 \times 10^{50}$\,erg of the original, unperturbed 
 star, 
 it indicates that the post-eruption stellar envelope remains somewhat inflated compared to a star in full thermal equilibrium.
 If this gives a sufficiently enhanced luminosity, it could lead to an extended epoch of further mass loss through a quasi-steady, super-Eddington, continuum-driven wind \citep{Quataert16,Owocki17}.
Detailed consideration of this will be left to future work.

\section{Energy Addition to Core }

\subsection{Adding half the star's binding energy to its core}

To explore how the ejection of mass depends on the location of the added energy, let us next consider a case with the same addition factor $f=0.5$, but now with the added energy distributed evenly in mass over the stellar {\em core}, with $M_{r} < 76 \Msun$.
Figure \ref{fig:fig7} shows key results for such a model.

The top panel plots contours of the mass  coordinate  $M_{r}$ vs. log radius and time over the full range out to the final simulation time, here just $600$\,ks.
Note that all the mass below $M_{r} < 98 \Msun$ now remains bound, but is excited into a pulsation with a gradual subsequent damping.

The middle panel shows that the Bernoulli energy undergoes a sharp jump from a shock that reaches from the core to near-surface after just $\sim13$\,ks.
The red contour for $B_{\rm n} = 0$ delineates the unbound mass above from the bound (oscillating) mass below.
Extensions to longer time shows that only about $\DMtot \approx 1.7 \Msun$ attains the positive net energy needed to escape, which is substantially lower than the $7.2 \Msun$ lost in the  standard $f=0.5$ case of energy addition to the stellar envelope.

The bottom panel plots the scaled density vs.\ similarity speed $r/t$.
Note that this now extends out to much higher speeds ($>$5000\,km\,s$^{-1}$) than in the fiducial case of envelope heating.
Indeed, at speeds above a ``breakout'' speed $ v_{\rm b} = 3500$\,km/s, the density no longer fits the usual exponential decline (shown by the dashed line in this semi-log plot), but rather follows a steep {\rm power-law} (dotted curve) defined by
\beq
\rho t^3 = d_{\rm p} \left ( \frac{1000 \, {\rm km \, s^{-1}}}{v} \right )^{9}  ~ ; ~ v> v_{\rm b} 
\, ,
\label{eq:dp}
\eeq
where the visual best-fit gives $d_{\rm p} \approx 2.5 \times 10^{9}$\,g\,cm$^{-3}$\,s$^3$.
Integrating this similarity form from the breakout speed $v_{\rm b}$ shows that the associated mass in this power-law tail is actually very small,
\beqa
 \Delta M_{\rm tail} 
 &=& 4 \pi \int_{v_{\rm b} t}^\infty \,r^2 \rho(r,t)  \, \textrm{d}r
 \nonumber
 \\
&=& 4 \pi d_{\rm p} \int_{v_{\rm b}}^\infty \,  v^2
\left ( 
\frac{
1000 \, {\rm km \, s^{-1} } }{v} 
\right )^9
 \, \textrm{d}v
  \nonumber
 \\
&\approx&  0.0014 \, \Msun
 \, .
 \label{eq:DMtail}
 \eeqa
This is only about $0.08$\% of the total ejecta mass $\DMtot \approx 1.7 \Msun$ (as estimated from the mass with a positive total energy at the final time $t_{\rm fin} = 600$\,ks).
Analogous integration for the kinetic energy shows that this power-law tail also represents only a small fraction, $\sim$2\%, of the total kinetic energy of the ejecta, $\DKEtot \approx 1.9 \times 10^{49}$\,erg.  The mean speed of the ejecta is ${\bar v} = 1055$\,km\,s$^{-1}$, again very near the stellar escape speed, $v_{\rm esc} = 1044$\,km\,s$^{-1}$.

The cumulative distribution of ejected mass fraction vs. scaled speed, defined as $m(x)$ in eqn.\ (\ref{eq:mxdef}) for the case of a purely exponential similarity, can be readily generalized to take account of this power-law tail for $x > x_{\rm b} \equiv v_{\rm b}/v_{\rm o}$,
\beqa
m(x,x_{\rm b}) &=& m(x) ~~~~~~~~~~~~~~~~~ ; ~~ x \le x_{\rm b}
\nonumber
\\
&=& m(x_{\rm b}) \left ( \frac{x_{\rm b}}{x} \right )^{6} ~~ ; ~~ x \ge x_{\rm b}
\, .
\label{eq:mxpowtail}
\eeqa

In summary, such a core heating model leads to initial, high-speed ejecta that has only a small fraction of the total ejected mass, which itself is much less than found in the standard envelope heating model with the same energy addition factor $f=0.5$.

\subsection{Adding twice the star's binding energy to its core}

To explore potential links to core-collapse SNe, let us next consider a model in which the energy added to the core is {\em twice}  the binding energy (i.e., $f=2.0$), and so could in principle completely disrupt the entire star.

Figure \ref{fig:fig8} shows results. As in figure \ref{fig:fig7}, the middle and bottom panels again show respectively the evolution of the Bernoulli energy, and the variation of scaled density with the similarity speed $r/t$.
In the former, the red contour for $B_{\rm n} =0$ shows that the parcels deep into the stellar core are developing a positive energy, which eventually leads to ejection of more than $87 \Msun$.
The latter again shows a strong power-law tail (dotted curve) at high similarity speed (now even exceeding 10,000\,km\,s$^{-1}$), with the same form as eqn. (\ref{eq:dp}), but now with a lower transition breakout speed, $v_{\rm b} = 2750$\,km\,s$^{-1}$.  
This and the higher  best-fit normalization ($d_{\rm p} \approx 1.34 \times 10^{11}$\,g\,cm$^{-3}$\,s$^3$) gives a higher tail mass $\Delta M_{\rm tail} \approx 0.33 \Msun$ and kinetic energy $\Delta K_{\rm tail}  \approx 7.4 \times 10^{49}$\,erg.
The corresponding fractions of total ejecta mass ($\sim$0.4\%) and kinetic energy ($\sim$11\%) are thus greater than for the  $f=0.5$ case, but still quite modest.

Indeed, it seems that the bulk of the mass ejection follows closely the similarity solution based on the exponential density decline in similarity speed. The top panel of figure \ref{fig:fig8} shows the mass distribution of the speed $v$ at the cited times. There is quite remarkably good agreement with the analytic form (\ref{eq:vMr}) (dashed curve), apart from the slight kink in the simulation results at the mass $M_r \approx 76 \Msun$ near the core-envelope boundary 
(marked by arrows in figure \ref{fig:fig8}).
This suggests that even the marked differences in stellar structure between core and envelope have only a relatively minor effect on the density and speed of the mass ejecta.
Thus the similarity forms based on the exponential scaling $\exp (-r/v_{\rm o} t)$ for density may apply quite robustly to eruptive mass loss from both core and envelope heating with energy addition factors of order unity.
For such cases,  gravity still plays a key role in regulating the mass ejecta properties, for example setting the mean speed to be of order the surface escape speed.

But these models with energy added in the core also lead to a prompt, high-speed ejecta from the sudden shock breakout through the stellar surface, with density following a power-law instead of an exponential decline in similarity speed.
Similar shock breakouts occur in core-collapse SNe \citep{Dessart10,Ro13,Ro17}, with the resulting ejecta also following the simple similarity form for the velocity, $v = r/t$.
Analysis of the mass and density distribution of such SN ejecta by
\citet{Matzner99} shows a rather complex dependence on the structure of the pre-SN star;
but to the extent that the ejecta density follows a similarity relation, \citet{Matzner99} fit it with a broken power law in $v=r/t$.

The appearance of a similar power-law tail in velocity for the core-heating cases here provides a potential bridge toward such scalings for SNe.
One may conjecture that further increases in the energy addition factor to large $f \gg 1$, as  occur in SNe, will lead to greater prominence of such a scale-free power-law.
As gravitational binding becomes increasingly overwhelmed, there should be a concomitant diminution in the exponential similarity form, since this enforces the role gravity through the characteristic speed $v_{\rm o}$,
 which sets the mean speed to near the stellar escape speed.
Detailed exploration of such transition from eruptive to explosive mass ejection is 
left for future work.

\begin{figure}
\includegraphics[scale=.67]{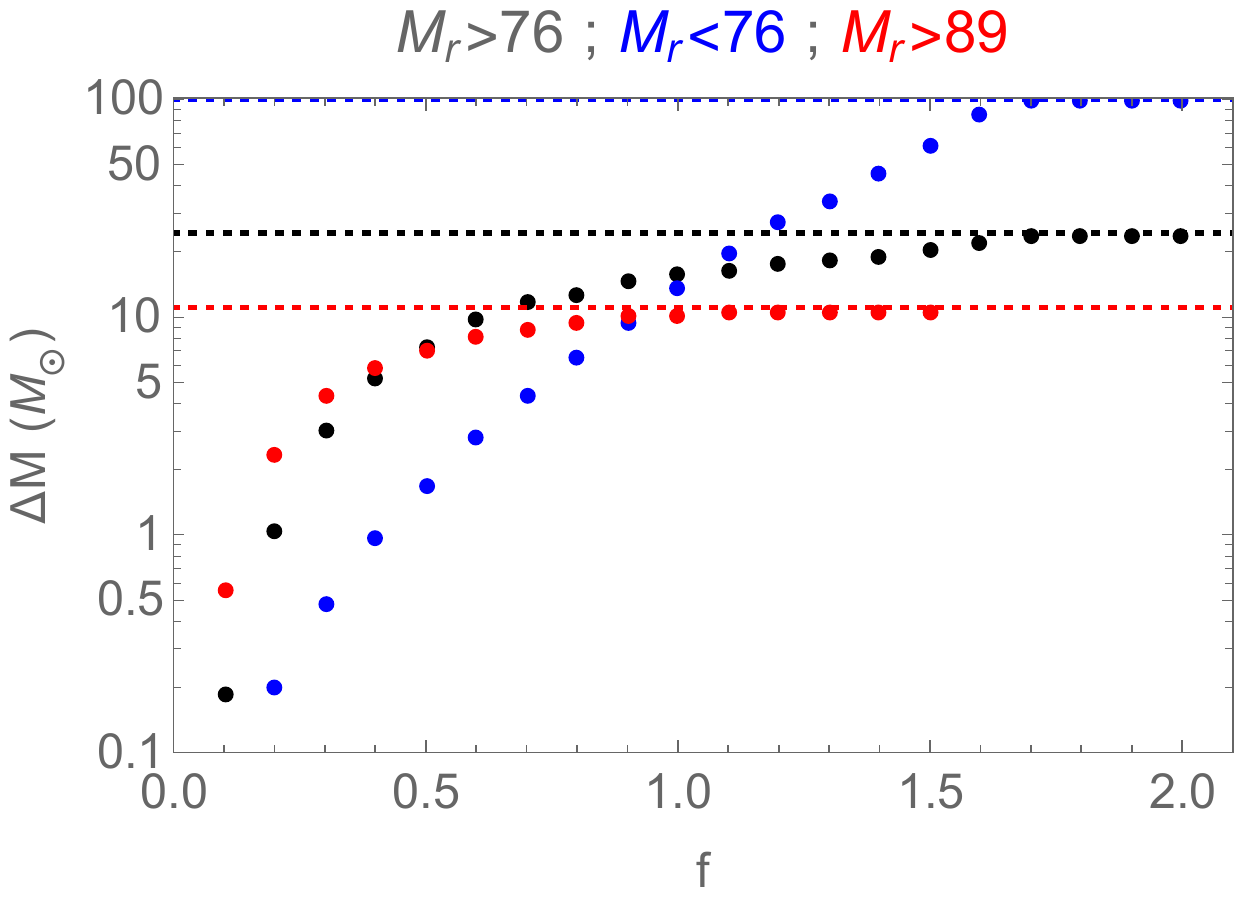}
\\
\\
\includegraphics[scale=.66]{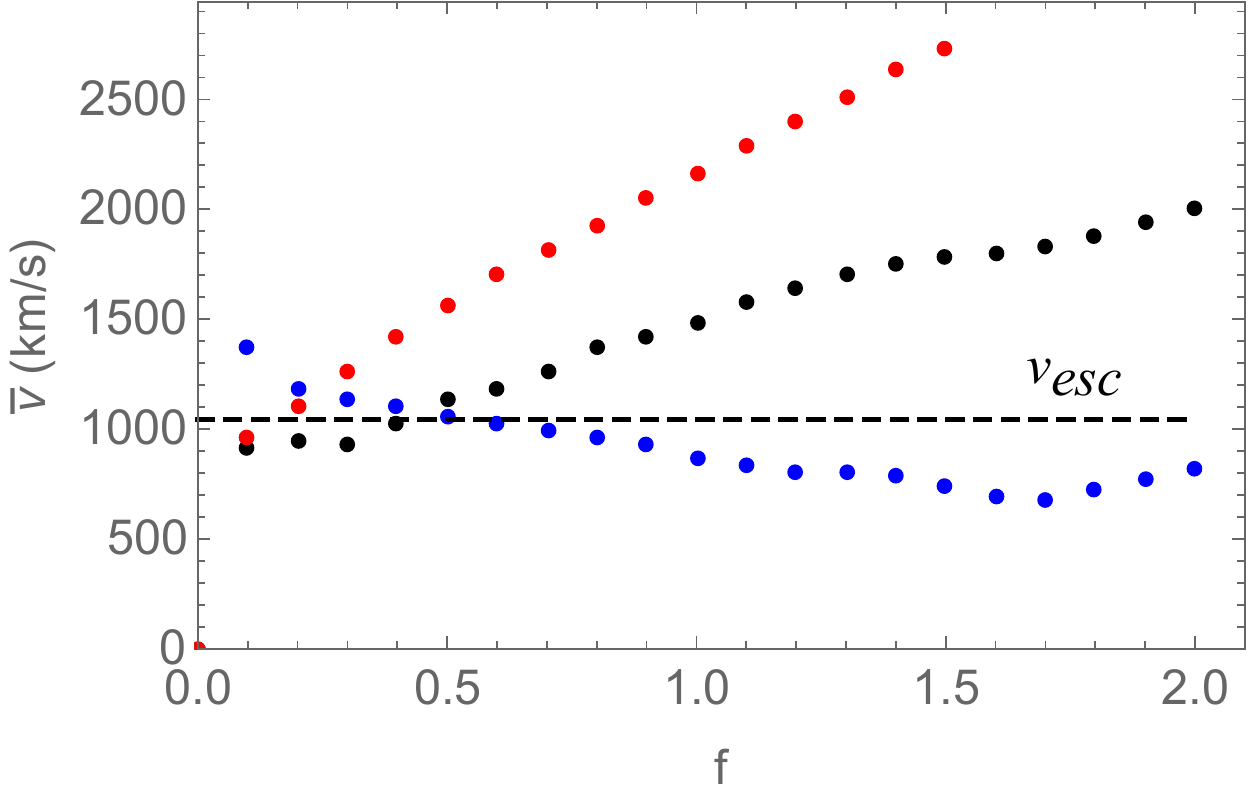}
\\
\\
\includegraphics[scale=.66]{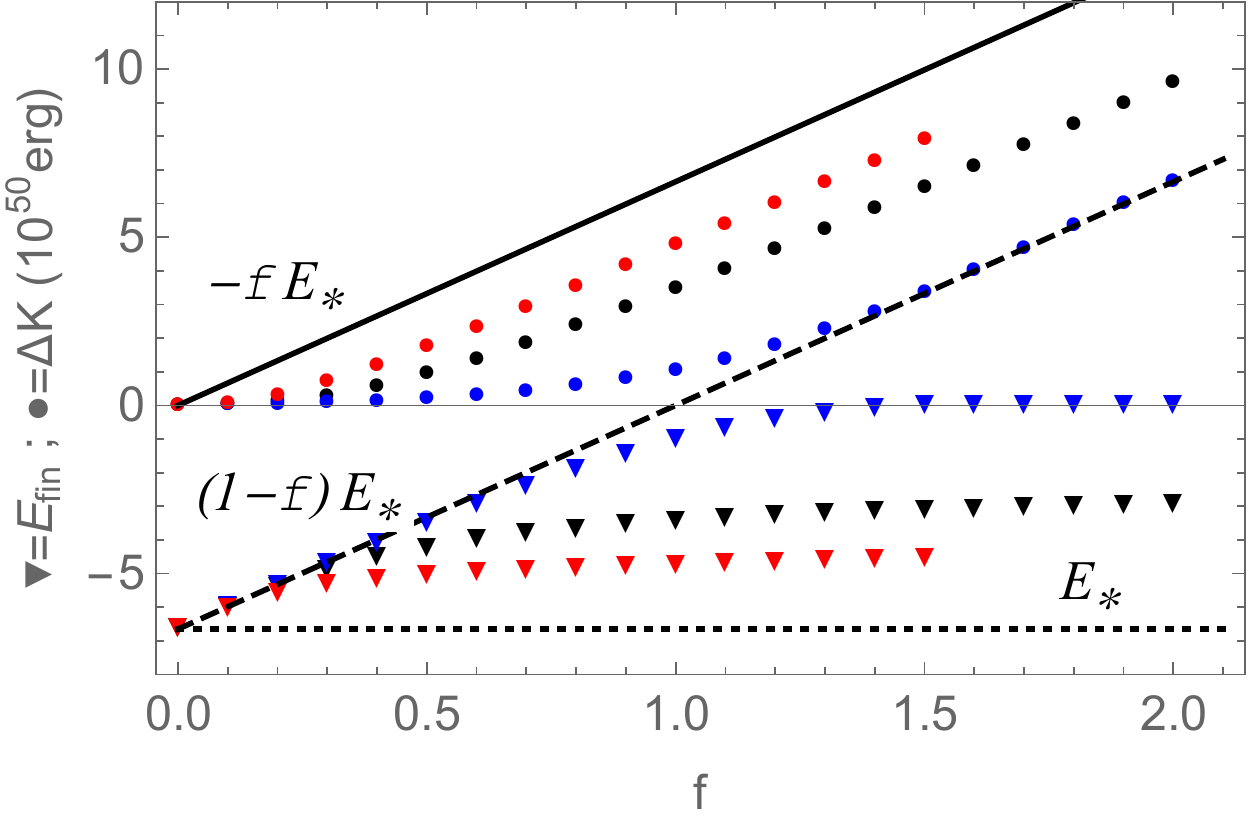}
\caption{
Key results from our parameter study, plotted vs. energy addition factor $f$, 
for models with energy addition in the core ($M_r < 76 \Msun$; blue points),
near the surface ($M_r > 89 \Msun$; red points),
and in the envelope just outside the core with ($M_r > 76 \Msun$; black points).
The top panel shows the total ejected mass, 
with the horizontal dashed lines for the mass above the onset of heating, representing an upper limit to ejected mass.
The middle panel compares the energy-averaged ejecta speeds
${\bar v}$ 
(defined in eqn.\ (\ref{eq:vbar})) with the surface escape speed  $v_{\rm esc} = \sqrt{2 G \Mstar/\Rstar} = 1044$\,km\,s$^{-1}$ (horizontal dashed line).
The bottom panel compares  the  kinetic energy of the ejecta ($\DKEtot$;  circles) and final stellar energy state ($\BEfin$;  triangles) with 
the binding energy of the unperturbed star ($\Ebind$; dotted line),
the initial stellar energy ($(1-f) \Ebind$; dashed line),  and the net added energy ($-f \Ebind$; solid line).
}
\label{fig:fig9}
\end{figure}

\section{Parameter study}

\subsection{Dependence of ejecta mass and energy on level and location of energy addition}

Let us next examine how the total mass and kinetic energy of the ejecta depend more broadly on the energy addition factor $f$, as well as on the location of this added energy.
For this we have
computed  a grid of models with energy factors ranging from $f=0.1$ up to $f=2.0$ 
and with three distinct locations for the energy addition, namely the 
core  ($M_r < 76 \Msun$), 
envelope  ($M_r > 76 \Msun$),
and near-surface ($M_r > 89 \Msun$).
Each model is computed to a final time (typically $\sim1$\,Ms) that is sufficiently long to show convergence for the mass coordinate 
with zero Bernoulli energy, $B_{\rm n} (M_e,t) \rightarrow 0$, giving then for the total ejected mass 
$\DMtot = \Mstar - M_e$.

As in the fiducial model (cf. eqn. \ref{eq:BEfin}),  
the total kinetic energy of the ejecta can be estimated from the drop in the energy of the star from the initial to final state,
\beq
\DKEtot 
\approx (1-f) \Ebind - \BEfin
\, ,
\label{eq:dketot}
\eeq
where the latter can be readily determined from the simulation.
 From eqn.\ (\ref{eq:vbar}), the total mass and kinetic energy of  the ejecta can then be used to compute its energy-averaged speed ${\bar v}$.

The top panel  of figure \ref{fig:fig9} shows that,
for all three locations of the energy deposition,
the ejected mass increases with $f$;
but at the largest $f \rightarrow 2.0$ they each approach a distinct asymptote, set by the total mass {\em above} the onset of heating, as indicated by the horizontal dotted lines.
In effect, even a large energy addition only ejects mass that lies above the deepest energy deposition.
For the core heating model, this means {\em all} the mass can be ejected, but for the envelope and surface heating cases, this is much more limited,
with asymptotic ejecta values of respectively $24 \Msun$ and $11 \Msun$, representing the entire mass of the heating region.


For most cases, the unbound mass is just a small fraction of the total mass, e.g. $<$10\% for the f$<0.5$ models, which is the relevant range for direct energy input by stellar mergers. 
This is consistent with the $<$8\% mass loss found in simulations of head-on collisions of high-mass stars
\citep{Glebbeek13},
though the details of the energy budget in those simulations are complex and thus still unclear.
Simulations of the dynamical phase of common envelope evolution 
\citep[e.g.,][]{Taam00,Passy12,Ricker12,Nandez16,Ohlmann16,Iaconi18}
also indicate inefficient mass ejection, but  until very recently \citep[see, e.g.,][]{Ricker18}, these have generally concentrated on low-mass stars, which have significantly different stellar structures and energy budgets.
In multi-D simulations the energy from mergers can take different forms like rotation or complex circulations not accounted for in the present idealized 1D simulations. 
But the breadth of this 1D parameter study can, with appropriate caution, provide a helpful general guide for interpreting the mass and energetics of ejecta in  more realistic multi-D  models.s

The{ middle panel of figure \ref{fig:fig9} shows the associated mean ejecta speeds ${\bar v}$, defined in eqn.\ (\ref{eq:vbar}).
For small energy addition factor $f$, the mean speeds for all cases are comparable to the surface escape speed of the unperturbed star, 
${\bar v} \approx v_{\rm esc}$, 
as shown by the horizontal dashed line.
This is reminiscent of the tendency of a steady stellar wind to reach a terminal speed that scales with the surface escape speed, $v_\infty \sim v_{\rm esc}$ \citep{Puls08,Owocki10}, reflecting the fact that the driving needed to lift material from the star tends also to do a comparable amount of work against inertia to propel the wind's kinetic energy.
In the present eruptive context, it means that as long as the mass of the heated region does not limit the ejected mass, there is a rough {\em equipartition} between gravitational and kinetic energy in the ejecta.
However, for envelope and near-surface energy deposition models, the saturation of the ejecta mass to that of the deposition region means that for higher energy cases $f>1$, the increased energy deposition leads to a higher specific kinetic energy, and thus higher ejecta speeds, ${\bar v} > v_{\rm esc}$.
An analogous increase in stellar wind speed occurs when the mass loss from the surface is ``choked'', e.g. by using a reduced density at the lower boundary \citep{Owocki10}.

The bottom panel of figure \ref{fig:fig9} shows that, 
for small $f$, only a small fraction of sthe added energy is imparted to the ejecta as kinetic energy, but for larger $f \rightarrow 2.0$, the ejecta kinetic energy increases in direct proportion to the added energy,
with an offset that increases with the depth of the energy addition.
The difference $-f \Ebind - \DKEtot$ represents energy left in the stellar envelope, leading to final net binding energy $\BEfin = (1-f) \Ebind - \DKEtot$ (cf.\ eqn.\ (\ref{eq:dketot})).
For the case of core energy deposition, we find $\BEfin  \rightarrow 0$ 
as $f \rightarrow 2.0$, showing that the entire star becomes disrupted.

For the cases with energy deposition in the envelope or near surface, there always remains a negative net binding energy for the retained mass below the heating region. 
To the extent that this is above the binding energy of the original, unperturbed star ($\Ebind = -6.65 \times 10^{50}$\,erg), it indicates the remaining star is likely to be left in a thermally excited state; if the associated luminosity exceeds the Eddington limit,  this could drive a strong, super-Eddington wind for up to a thermal relaxation timescale,  resulting in a post-eruption mass loss that could cumulatively rival that from the initial ejecta.
This is an important topic for future investigation.

In summary, for modest energy factors $f<1$ the ejecta mass increases with increasing $f$, with ejecta speed comparable to the stellar escape speed.
For large factors $f>1$, the ejection mass for the envelope and near-surface heating model is limited to the mass in those regions, with the speed and kinetic energy then increasing with increasing energy factor.
By contrast, adding such large energy in the stellar core can completely disrupt the star, ejecting up to the entire stellar mass with average speed near the surface escape speed.

Appendix B develops a simple fitting function for the $f$-dependence of the mass ejecta (scaled by the mass above the bottom of the heating region, i.e. $\DMtot/M_{\rm bot}$),  while figure \ref{fig:figB1} compares the visual best fits with simulation results for each of the three locations for the energy addition.

\subsection{Applicability of similarity forms and relation to SNe}

As noted at the end of section 5.2,
the extension of eruption simulations to higher energy factors above unity, and the inclusion of models with the energy addition in the stellar core,
provides a potential bridge to the properties of core-collapse supernovae (SNe), for which $f \gg 1$, with the energy concentrated near the center of the enriched core.
A full discussion is beyond the scope of the current work aimed at modeling eruptive mass loss,
for which the assumed pre-eruptive interior and core structure is much different than for SNe;
but in this context, let us briefly consider the nature and applicability of similarity relations in SNe vs.\ the eruption models here.

While SN ejecta are  commonly modeled in terms of a similarity form for the velocity $v = r/t$,
the analysis by \citet{Matzner99} of their mass and density distribution shows a rather complex dependence on the structure of the pre-SN star.
But to the extent that ejecta density follows a similarity relation, \citet{Matzner99} fit it with a broken {\em power law} in $v=r/t$.
This is contrasts with the {\em exponential} density decline found for our eruptive models, $\rho \sim e^{-r/v_{\rm o} t}$, with the characteristic velocity $v_{\rm o}$ playing a key role in the resulting scalings of the ejecta mass and kinetic energy, as given by eqns. (\ref{eq:DMtot}) and (\ref{eq:DKEtot}).

Analysis of our extended parameter study indicates that for energy addition factors $f \lesssim 1$, both the model series with surface and envelope energy addition also show this exponential decline in density, with the characteristic velocity following roughly the $v_{\rm o} \approx {\bar v}/\sqrt{12}$ scaling given in eqn.\ (\ref{eq:vbar}).
The mass distribution of velocity likewise follows the similarity form given by eqn. (\ref{eq:vMr}).

For such cases with higher energy addition, $f>1$, the time evolution can be more complex, and not always well fit with this simple similarity form for density.
The strong energy deposition in the envelope or near the surface excites oscillations in the underlying core, leading to repeated outward shock waves (similar to the reflected shock in the $f=0.5$ envelope model; see figure \ref{fig:fig2}), and so an extended time interval before the dynamical evolution settles into a simple similarity form.

For the models with core energy addition there are no such reflected shocks, so we do generally find the evolution approaches a similarity form for the full range of energy factors $0.1 < f < 2.0$.
Specifically, at moderate $v=r/t \lesssim 3000$\,km\,s$^{-1}$, the density agains shows an exponential decline $e^{-r/v_{\rm o} t}$, but then shifts to more of a {\em power-law} for large speeds  $v \gtrsim 3000$\,km\,s$^{-1}$.
We interpret this as resulting from the greater strengthening of the shock over the envelope propagation from its origins in the heated core, with then the surface layers expelled at high speed in a way similar to shock breakout in SNe
\citep{Dessart10,Ro13,Ro17}
But for the greater mass in the subsurface layers, the non-negligible retarding effect of the gravity leads to an exponential density cutoff, with the characteristic velocity $v_{\rm o}$ reflecting the competition between gravity and energy addition that sets the bifurcation between bound and ejected mass.
In contrast, the models with energy added to the envelope or near-surface don't have this extended shock growth, and so have little or no material ejected with the high speeds associated with this power-law tail in similarity variable $v=r/t$.

There is also a potential analogy between such eruptions and possible {\em failed} explosions from core-collapse
\citep{Dessart10,Lovegrove13,Coughlin18a}. 
In the latter, the energy from collapse escapes by neutrino emission, with insufficient coupling to the stellar envelope to induce an explosion;
but the mass lost in neutrinos and the  associated reduced gravitational binding lead to a weak, low-speed ejection of some of the envelope mass
\citep{Lovegrove13} in a way that is in some respects similar to the eruptions from the mass-proportional addition of energy in the models here.

We thus see that the core heating models here do indeed suggest some potential links between such eruptions and both SNe explosions and failed explosions, but we leave a more complete examination of this to future work,
which should include examination of the potential role of fundamental differences in the pre-explosion stellar structure.

\section{Concluding summary and future outlook}

Motivated by the eruptive mass loss inferred from LBV stars, particularly the 1840's great eruption of $\eta$\,Carinae, this paper uses time-dependent hydrodynamical simulations to examine the mass ejection that results from impulsive energy addition to the interior of a very massive star.
This contrasts with previous models of steady-state energy addition that results in a continuum-driven super-Eddington stellar wind \citep{Quataert16,Owocki17}.
Moreover, unlike supernova explosion models -- for which the extreme energy addition in the core leads to complete, high-speed ejection of the overlying stellar material --, the energy addition here is at an order-unity factor of the stellar binding energy, and so results in ejection of only some outer fraction of the stellar mass, with the rest falling back onto an expanded, though still bound, surviving star.

As in SNe, the velocity of the expanding ejecta closely follows a similarity relation in radius and time, $v \approx r/t$.
But a further, key result is that the density of this expansion closely fits a declining exponential $\rho \sim e^{-r/v_{\rm o} t}$ (see figure \ref{fig:fig5} and eqn. (\ref{eq:rhofit})), with now a characteristic velocity scale $v_{\rm o}$ that
enforces the effect of gravitational binding in limiting the ejected mass to the outermost material.
This contrasts with the scale-free, power-law similarity scalings inferred for SNe \citep{Matzner99}, and to our knowledge  is a distinctly novel property
for such eruptive mass loss with a bifurcation between ejected and bound mass (see, however \citet{Dwarkadas98}).

Integration of this exponential similarity form for density yields analytic scalings for the ejected mass and kinetic energy (eqns.\ (\ref{eq:DMtot}) and (\ref{eq:DKEtot})), with an associated analytic form for the mass distribution of the velocity (see figure \ref{fig:fig6} and eqn.\ (\ref{eq:vMr})).
An associated scaling for the temperature (eqn.\ (\ref{eq:Tfit})) should prove useful for future analysis of the observational signatures of such eruption models.

Our broader parameter study shows how the ejecta mass and kinetic energy depend on the level and location of the energy 
addition\footnote{It likely also will  depend on the stellar structure, but since we here have fixed the stellar model, quantifying this will require future investigation; see below.}.
For energy addition less than the stellar binding energy ($f<1$), the ejected mass increases with increasing $f$, with an average flow speed generally near  the surface escape speed, ${\bar v} \approx v_{\rm esc}$.
For energy addition exceeding the binding energy by up to a factor $f=2$, results depend on the location of the energy deposition.
For addition to the envelope or near-surface, the ejected mass is limited to the mass of the heated region, with the excess energy thus increasing the ejecta's specific energy and thus flow speed.
For addition in the stellar core, such a large energy factor can lead to ejection of nearly all the stellar mass, but with slower speed. 
This provides a potential bridge between the present eruption models and full SN explosions that warrants further investigation.

In the present context of exploring the nature of eLBV stars, the dynamical simulations here have assumed a very massive star at an intermediate main-sequence evolutionary phase.  Future work will be needed to explore how the specifics of mass ejection depend on stellar mass and evolutionary state. For example, in their exploration of a binary merger scenario for SN1987A, \citet{Morris07,Morris09} analyzed how this could lead to anisotropic ejection of the extended envelope of a red supergiant. The low binding energy of this envelope facilitates its ejection, but the ejected mass is much lower than in the eLBV scenarios explored here. Overall, the form, speed and mass of the ejecta should depend on both the details of the energy and angular momentum addition, and the mass and evolutionary state of the merging stars. It will be of interest to explore how  some of the key findings of the present study, e.g. the exponential similarity of the ejecta density, may extend to such a broader range of mass eruptions.

In general, the hydrodynamical simulations and similarity relations presented here provide important new insights into the properties of the eruptive mass loss that can arise from impulsive energy addition to various regions of a massive-star interior.
The scales of  the ejected mass, over a range from $< 0.1 \Msun$ to $> 10 \Msun$, and of the ejecta speeds, from a few hundred to several thousand km\,s$^{-1}$, are comparable to what's inferred from eLBV's \citep{Smith11}.
Moreover, a small amount of material is ejected at a speed much higher than the average, providing a possible rationale for the high-speed ejecta inferred by \citet{Smith18b} for  $\eta$\,Carinae \citep[see also][]{Smith06a,Smith08a,Smith13}.

But clearly the 1D, adiabatic models here -- with an assumed impulsive energy addition of unspecified physical origin -- represent only one further step toward understanding eruptive mass loss from massive stars.
Future improvements should include a treatment of radiative transport, particularly for the near-surface layers that will set the observational signatures of the eruptions.
Any more detailed comparison with specific eLBV's, most notably $\eta$\,Carinae, will also require significant extensions, to include a specific mechanism for the energy addition, and to identify its level and location, and the degree to which it can be characterized as impulsive, steady, or, most likely, something intermediate between these idealizations.
In particular, in the case of a binary merger model for  $\eta$\,Carinae, there is a need for 2-D and 3-D dynamical models that account for the role of the binary orbital angular momentum in spinning up the merged star and  centrifugally ejecting mass, while also accounting for the energy deposition from merger of the separate stellar cores.

\section*{Acknowledgments}
We acknowledge Nathan Smith for helpful discussions and many useful comments and suggestions on an early draft.
We thank Chris Matzner for pointing out to us previous findings of exponential similarity scalings for density in SNIa explosions.
We also thank Lorne Nelson for sharing computing facilities,
and the referee, Alex de Koter, for constructive criticisms that helped improve the final manuscript.
The computations here were partially carried out on facilities managed by Calcul Qu\'ebec and Compute Canada.
RH was supported by the JSPS Overseas Research Fellowship No.29-514 
and a grant from the Hayakawa Satio Fund awarded by the Astronomical Society of Japan.
SPO acknowledges support of a visiting scholar grant from the Royal Society.
This work has also been supported by a Humboldt Research Award to Ph.P. at the University of Bonn, 
and by the Oxford Hintze Centre for Astrophysical Surveys, which is funded through generous support from the Hintze Family Charitable Foundation.
Many of the ideas for this study grew out of discussions during a massive-star research program at the Kavli Institute for Theoretical Physics, 
and as such was supported in part by the National Science Foundation under Grant No. NSF PHY-1125915. 

\bibliographystyle{mn2e}
\bibliography{OwockiS}

\appendix

\newpage

\section{Similarity relations for internal energy and temperature}

Let us also examine the time and space variation of the temperature for ejected material 
within this idealized model of adiabatic expansion.
To do this, we first consider the variation of total internal energy $E= 3 P_{\rm rad} + (3/2) P_{\rm gas}$, as given by eqn.\ (\ref{eq:Econ}).
Applying a similarity form for the velocity $v=r/t$, this can be written as
\beq
\frac{DE}{Dt}  
= - 3 \frac{E+P}{t} \, 
\, .
\label{eq:Econsim}
\eeq
Defining $\beta \equiv P_{\rm gas}/{P}$, we can recast this in the form,
\beq
\delta \equiv
- \frac{D \ln E}{ D \ln t}  
=  3  \left ( 1  + \frac{P}{E} \right )
=  3 + \frac{1 }{1 - \beta/2} 
\, .
\label{eq:DlnEdlnt}
\eeq
This implies that, if $\beta$ were constant, the internal energy of mass parcels with fixed $M_r$ would decline as a simple power law in time,
$E \sim 1/t^{\delta}$.
Note in particular that the power index ranges from $\delta = 4$ for the radiation-dominated ($\beta = 0$) case, to $\delta = 5$ in the gas-dominated ($\beta = 1$) limit, with $\delta = 4.33$ for the special case $\beta=1/2$ with equal gas and radiation pressure.

\begin{figure}
\begin{center}
\includegraphics[scale=.65]{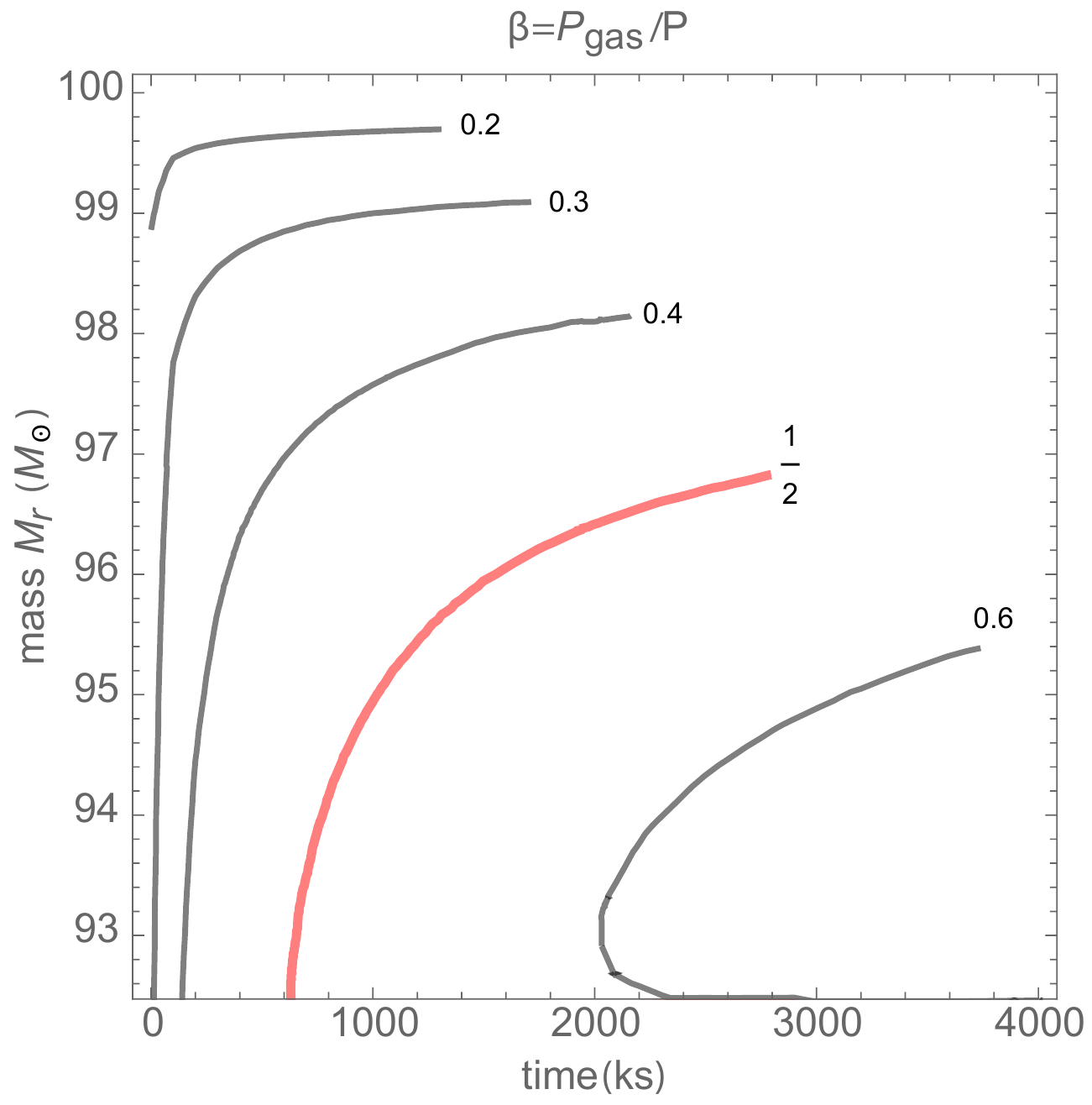}
\caption{
For the fiducial model with energy-addition factor $f=0.5$ applied in the stellar envelope,
contours of gas pressure fraction $\beta$ plotted vs.\ mass and time, showing that most of ejected material has comparable gas and radiation pressure, with $\beta = 0.5$ highlighted in red.
The termination of the contours in the upper right represents the outer boundary of the computational domain at $R_{\rm max}=4000 \Rsun$.
}
\label{fig:figA1}
\end{center}
\end{figure}

\begin{figure}
\begin{center}
\includegraphics[scale=.5]{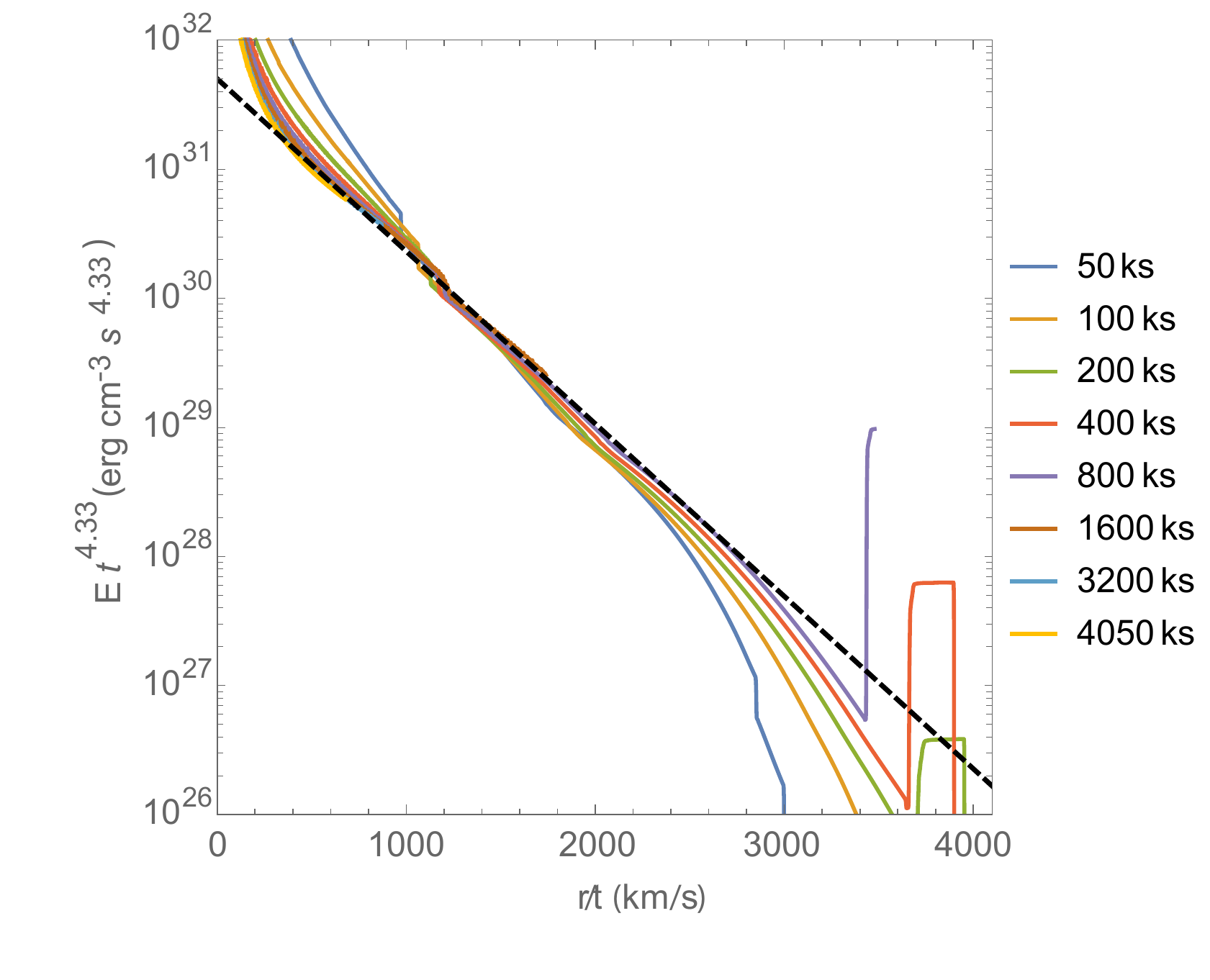}
\caption{
Plots of the internal energy vs.\ similarity variable $r/t$ for the cited time snapshots in the fiducial $f=0.5$ model. 
The ordinate shows the
time-scaled internal energy $E t^\delta$ for an exponent $\delta = 4.33$ appropriate for the equipartition gas pressure fraction $\beta \approx 0.5 $, as found in figure \ref{fig:figA1} to hold approximately through much of the ejecta.
The dashed line shows the visually best-fit similarity form, which has a slope consistent with the cutoff speed $v_{\rm o} = 325$\,km s$^{-1}$ found also for density in figure \ref{fig:fig5}.
}
\label{fig:figA2}
\end{center}
\end{figure}

For the fiducial model with energy-addition factor $f=0.5$ applied in the stellar envelope, figure \ref{fig:figA1} shows a contour plot of $\beta$ as a function of mass $M_r$ and time $t$. The heavy red line shows the equipartition value $\beta=1/2$, while the other contours show that for most of the ejecta, $\beta$ remains within about $\pm 0.2$ of this equipartion value.

Figure \ref{fig:figA2} plots time snapshots of the internal energy scaled by time as $E t^{4.33}$, i.e. with the exponent $\delta=4.33$ appropriate for $\beta \approx 0.5$.
The dashed line represents the best-fit linear trend, corresponding again in this semi-log plot to an exponential decline $e^{-r/v_{\rm o} t}$, with $v_{\rm o} \approx $ 325 km\,s$^{-1}$.
The total energy thus follows a similarity relation of the form,
\beq
E(r,t) = \frac{E_{\rm o} t_{\rm o}^{4.33}}{t^{4.33}} e^{-r/v_{\rm o} t}
\, ,
\label{eq:Esim}
\eeq
where the proportionality constant has the CGS value $E_{\rm o} t_{\rm o}^{4.33} \approx 6. \times 10^{31}$ erg cm$^{-3}$ s$^{4.33}$.

If we also apply the approximation $\beta \approx 1/2$ in the expression for total energy, we find 
\beq
E = 3 P_{\rm rad} + \frac{3}{2} P_{\rm gas} \approx \frac{9}{2} P_{\rm rad} = \frac{3}{2} a_{\rm rad} T^{4}
\, .
\label{eq:Ebetahallf}
\eeq
The overall similarity fit for the temperature thus takes the approximate form,
\beq
T(r,t) \approx \frac{ T_{\rm o} t_{\rm o}^{1.08}}{t^{1.08}} \, e^{-r/4 v_{\rm o} t}
\, ,
\label{eq:Tfit}
\eeq
where the scaling constant 
\beq
T_{\rm o} t_{\rm o} ^{1.08} = \left ( \frac{E_{\rm o} t_{\rm o}^{4.33}}{\frac{3}{2} a_{\rm rad}} \right )^{1/4} \approx 2.7 \times 10^{11} \,{\rm K \, s^{1.08}}
\, .
\label{eq:Totocon}
\eeq

These similarity relations for temperature, density, and velocity can be used, together with a prescription for the opacity, to model observational properties of the eruption, such as the light curve, or even the evolving spectrum.
But reliable results will also require inclusion of a radiation transport and the associated radiative losses near the expanding photosphere.
This will be a topic for future work.

\section{Fits to ejecta mass vs.\ energy fraction}

The parameter study in section 6 shows that the mass ejected at high energy-addition factor $f>1$ is limited to the mass above the bottom of the heating region, $M_{\rm bot}$.
But in the opposite limit of small energy factor, $f \ll 1,$ the mass ejection scales roughly as a power in $f$.
Let us thus define a general fitting function for the scaled mass ejection that bridges these two limits,
\beq
\frac{\DMtot}{M_{\rm bot}} (f) = 
\left [ 1+ \left ( \frac{f_{\rm b}}{f} \right )^{2n}  \right]^{-1/2}
\, ,
\label{eq:dmfit}
\eeq
where the parameter $f_{\rm b}$ represents the transition from the power-law form $\DMtot \sim f^{n}$ at low $f  \ll f_{\rm b}$,
to the saturation $\DMtot  \rightarrow M_{\rm bot}$ at large $f > f_{\rm b}$.
For the parameter values in Table \ref{table:B1}, figure \ref{fig:figB1} compares this fitting function to the mass ejection results for each of the three locations of energy addition.

\begin{figure}
\begin{center}
\includegraphics[scale=.65]{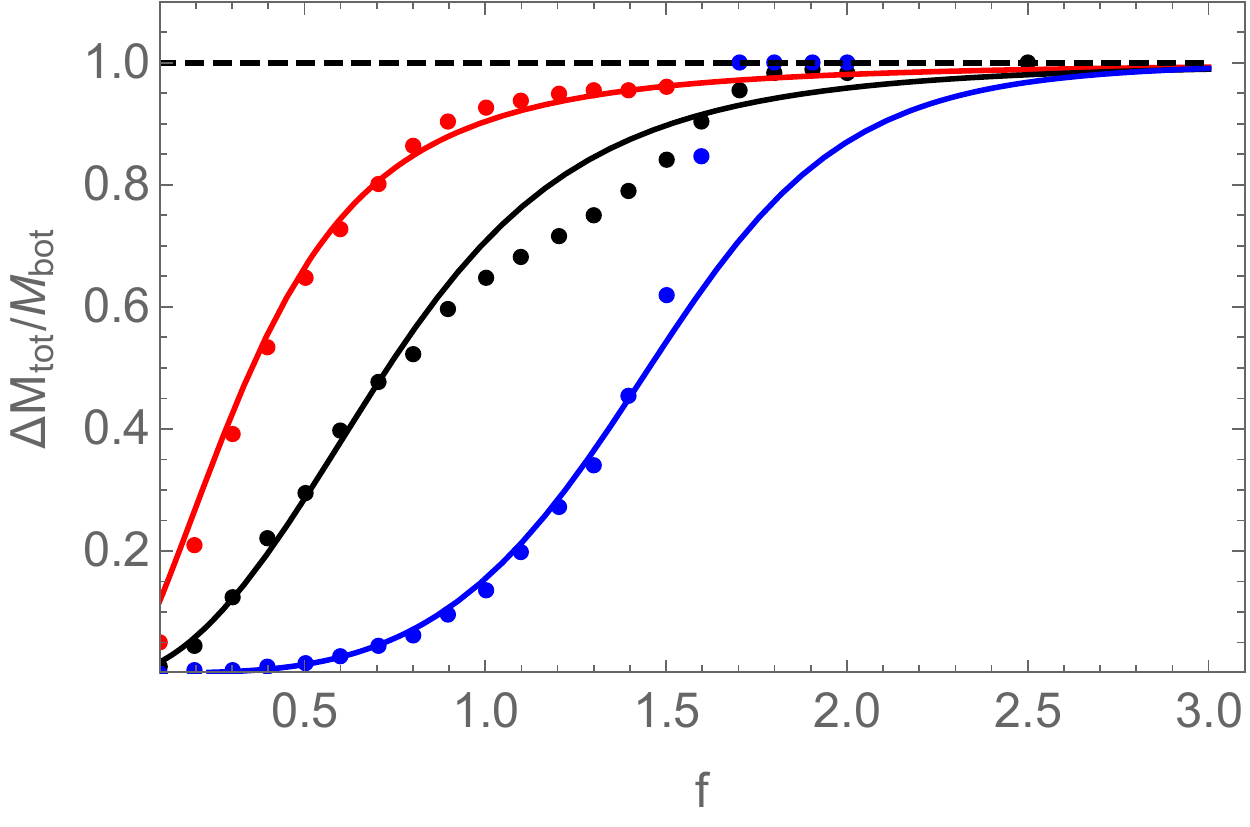}
\caption{For energy addition to the core, envelope, and near-surface (blue, black, and red), the mass ejection scaled by the mass above at the bottom of the heating region ($\DMtot/M_{\rm bot}$) vs. the energy addition fraction $f$.
The lines represents visual best fits from the fitting function (\ref{eq:dmfit}), with parameters given in Table \ref{table:B1}.
}
\label{fig:figB1}
\end{center}
\end{figure}

\begin{table}
\centering
\caption{
For the solid curves fits to the mass data in figure \ref{fig:figB1}, parameters for the fitting function (\ref{eq:dmfit}),
for energy addition near the surface, in the envelope, and in the core.}
\begin{tabular}{l c c c}
Heating region & $M_{\rm bot}$ & $f_{\rm b} $ & $n$ \\
\hline 
$M_{r} >  89 \Msun$ &$ 11 \Msun$ &0.55 &  1.25 \\
$M_{r} >  76 \Msun$ & $24 \Msun$ &1.00 &  1.75 \\
$M_{r} <  76 \Msun$ & $  100  \Msun$ &1.70 &  3.50 \\
[0.5ex]
\hline
\end{tabular}
\label{table:B1}
\end{table}

\end{document}